# An Efficient Bayesian Experimental Calibration of Dynamic Thermal Models

L. Raillon, C. Ghiaus
loic.raillon1@insa-lyon.fr, christian.ghiaus@insa-lyon.fr
Univ Lyon, CNRS, INSA-Lyon, Université Claude Bernard Lyon 1, CETHIL UMR5008, F-69621, Villeurbanne, France

**Abstract**

Experimental calibration of dynamic thermal models is required for model predictive control and characterization of building energy performance. In these applications, the uncertainty assessment of the parameter estimates is decisive; this is why a Bayesian calibration procedure (selection, calibration and validation) is presented. The calibration is based on an improved *Metropolis-Hastings* algorithm suitable for linear and Gaussian state-space models. The procedure, illustrated on a real house experiment, shows that the algorithm is more robust to initial conditions than a maximum likelihood optimization with a quasi-Newton algorithm. Furthermore, when the data are not informative enough, the use of prior distributions helps to regularize the problem.

**Nomenclature**
**Notations**

| | |
|---|---|
| $x, y, z$ | Scalars |
| $\mathbf{x}, \mathbf{y}, \mathbf{z}$ | Vectors |
| $\mathbf{A}, \mathbf{B}, \mathbf{C}$ | Matrices |
| $\mathbb{R}^q$ | Space of dimension $q$ |

**Notational conventions**

| | |
|---|---|
| $\mathbf{A}^{\mathrm{T}}$ | Matrix transpose |
| $\mathbf{A}^{-1}$ | Matrix inverse |
| $\mathbf{A}^{-1/2}$ | $\left(\mathbf{A}^{1/2}\right)^{-1}$ |
| $\mathbf{A}^{-\mathrm{T}/2}$ | $\left(\mathbf{A}^{-1/2}\right)^{\mathrm{T}}$ |
| $\det(\mathbf{A})$ | Determinant of the matrix $\mathbf{A}$ |
| $\mathrm{tr}(\mathbf{A})$ | Trace of the matrix $\mathbf{A}$ |
| $\dot{\mathbf{x}}$ | Time derivative of vector $\mathbf{x}$ |
| $\partial \mathbf{x}/\partial \theta_i$ | Partial derivative of $\mathbf{x}$ with respect to $\theta_i$ |
| $\mathrm{diag}(a_1, a_2, \ldots, a_N)$ | Diagonal matrix with diagonal values $a_1, a_2, \ldots, a_N$ |
| $\mathbb{E}[\cdot]$ | Expected value |
| $p(\mathbf{x})$ | Probability density function (pdf) of a random variable $\mathbf{x}$ |
| $p(\mathbf{x}|\mathbf{y})$ | Conditional pdf of vector $\mathbf{x}$ given vector $\mathbf{y}$ |
| $\mathbf{x} \sim p(\mathbf{x})$ | Random variable $\mathbf{x}$ with probability distribution $p(\mathbf{x})$ |
| $\propto$ | Proportional |
| $\approx$ | Approximately equal |
| $\mathbf{x}_{1:N}$ | Set of values $\mathbf{x} = [\mathbf{x}_1, \mathbf{x}_2, \ldots, \mathbf{x}_N]$ |

1. Introduction

The existing methods for characterizing building energy performance and energy saving provided by retrofitting are not relevant (Turner & Frankel 2008, De Wilde 2014). The energy performance estimation of buildings and associated systems must be independent of weather conditions and user behavior. From this assessment, the Efficiency Valuation Organization has developed the International Performance Measurement and Verification Protocol (IPMVP) (EVO 2014). The idea is to construct a physical model which characterizes the building intrinsic thermal dynamic and relate inputs to outputs measured on-site. Hence, the gap between the energy use given by the pre-retrofit and post-retrofit models represents the energy gained by the refurbishment.

Minimizing heat losses from buildings is the most obvious solution to reduce the heating and cooling demand but the efficiency and sustainability of the energy chain, from the production to the HVAC systems, must also be improved. Nowadays, the dominant paradigm is that the energy sources need to respond to all requests at any moment. The complexity of this strategy will be augmented with the increasing of the share of renewable energy sources in the energy mix. Therefore, supply and demand must become more flexible by using demand response mechanisms and energy storage (European Commission 2016). In order to adapt the demand to the production, the energy demand must be known. Physical models characterizing the thermal dynamic of buildings associated with model predictive control can be used to forecast the energy demand while maintaining indoor comfort (Hazyuk et al. 2012a, Hazyuk et al. 2012b, Ghiaus & Hazyuk 2010).

Two important societal needs are identified, the estimation of building energy demand and the estimation of energy savings brought by energy conservation measures. These two societal needs have the same scientific deadlock: the experimental estimation of the physical parameters of the dynamic thermal behavior of buildings.

Such models can be obtained considering the energy balance between buildings and their surroundings and energy balance in buildings can be modelled by using thermal networks (Naveros 2016, Ghiaus 2013). Stochastic state-space models are obtained by first transforming thermal networks in deterministic state-space, and then noise terms are added to represent the deviations between the differential algebraic equations and the true variations of the states. Stochastic state-space models relate inputs to outputs, where the dynamic of the states is given by the parameters; hence by knowing inputs and parameters, the output of the system can be simulated. However, the direct problem requires to know the parameters, so the inverse problem of parameter estimation must be solved first. The interest in parameter estimation for dynamic thermal models is not new (Nielsen & Nielsen 1984) and experiments of various scales have been used to test the validity of different approaches (Bloem 1994, Baker & van Dijk 2008, Jiménez 2014).

It is essential when making prediction or decision from an identified model to assess the uncertainties in the parameter estimates; it should be done by taking all the information available. From this perspective, Bayesian estimation gets more and more consideration, for instance, in quantification of energy saving from retrofit (Heo et al. 2012, Heo et al. 2015, Tian et al. 2016 and Li et al. 2016), in calibration of energy models (Chong & Lam 2015, Chong & Poh Lam 2017, Zayane 2011), in estimation of thermal characteristic of a wall (Berger et al. 2016) and in estimation of heating consumption (Kristensen et al. 2017). Bayesian methods, such as the *Metropolis-Hastings* (MH), are usually employed for low-dimensional problems because by increasing the number of parameters it becomes more and more difficult to tune properly the algorithms.

The paper compare the three phases of an experimental model identification (selection, calibration and validation) from a Bayesian and frequentist point of view. More precisely, the paper treats the problem of parameter estimation of dynamic thermal models when the model structure is known. First, the choice of a model structure is discussed and then an implementation of the second-order Metropolis-Hastings for linear and Gaussian state-space models is proposed. Tools and guidelines for tuning and diagnosing the algorithm are also presented. Next, different criteria are presented to assess the performance of models and to guide the selection of a model structure in agreement with data. The whole procedure is tested on a real test case where the differences between the Bayesian and the frequentist approach are illustrated.

2. Twin houses experiment

Twin houses are a real outdoor experiment conducted by the Fraunhofer Institute near Munich during April and May 2014. It is an unoccupied single family house (twin house O5) with 100 m$^2$ ground floor, a cellar and an attic space; a full description of the house and the experiment is given by Strachan et al. (2016). The experiment was designed such that the south zone (Figure 1) has only two boundary conditions: the external temperature and the adjacent spaces (cellar, attic, north zone). The adjacent spaces were held at 22 °C with blinds closed to reduce the chance of overheating and the doors separating the north and south zone were sealed off. The electric heaters on the south zone were synchronized on a Randomly Ordered Logarithmic Binary Sequence (ROLBS) to maintain a similar temperature in the different rooms (800 W in the living room, 500 W in the south bedroom and 500 W in the bathroom). The ROLBS signal was designed for three reasons: maximize the temperature difference with the boundary conditions, excite the range of time constant from 1 hour to 90 hours and decorrelate the heating signal with the solar radiation. The heaters are lightweight, with a time response estimated around 1 or 2 minutes by the

Fraunhofer Institute (Strachan et al. 2016) and a split coefficient between convective and radiative heat gains of 70/30 %. Mechanical ventilation was set to supply a volume flow rate of 60 m$^3$/h into the living room and extract 30 m$^3$/h in the bathroom and the south bedroom.

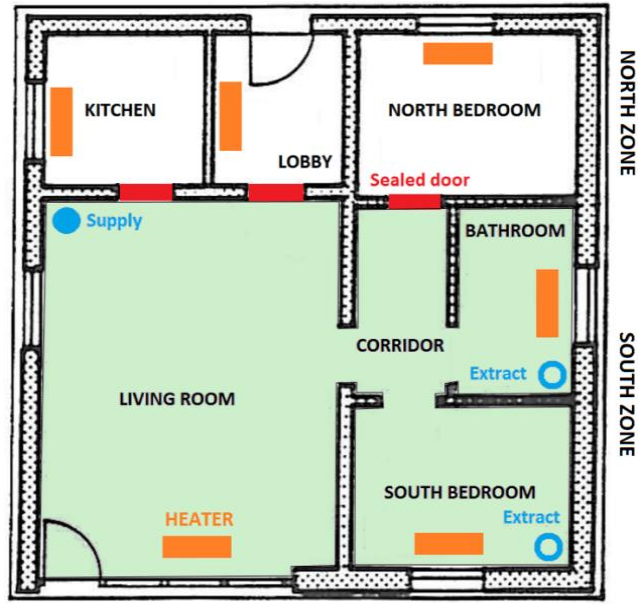

*Figure 1: Layout of the twin house O5*

The stratification of the air in each room of the south zone is measured with temperature sensors at 10 cm, 110 cm and 170 cm from the ground. This level of accuracy is not required to fit a simple dynamic thermal model; therefore the temperature of the rooms are chosen as the average of the three heights and the south zone temperature $T_s$(°C) is a weighted average of the spaces (living room, south bedroom, corridor and bathroom) by their respective surfaces. The boundary temperature $T_n$(°C) is also a weighted average of the temperatures in the different spaces (kitchen, lobby and north bedroom). A weather station near the house provides the outside air temperature $T_o$(°C) and the global solar irradiance measured on a horizontal surface $\dot{Q}_{gh}$(W/m$^2$). The heating in the south zone is done by three electric heaters; therefore $\dot{Q}_h$ (W) represents the total heat input injected.

Buildings are dynamic systems which may be modelled by using thermal networks from where state-space models can be deduced (Ghiaus 2013). This procedure is used to find a suitable model of the twin house where the south zone (in green on Figure 1) is considered as the main thermal zone and the north zone as an adjacent space.

3. Experimental calibration process

The experimental calibration process is decomposed in 3 phases: selection, calibration and validation (Ljung 2002). First, a set of likely model structures characterized by unknown physical parameter vector $\boldsymbol{\theta}$ is chosen based on a-priori knowledge (physics, experiment details, etc). Then, the calibration assesses how these model structures relate to observed data and to physical considerations. The calibration consists of finding a set of parameters which best represents the input-output relationship of a model through observed data, $\mathbf{u}_{1:k} = [\mathbf{u}_1, \mathbf{u}_2, \dots, \mathbf{u}_k]$ and $\mathbf{y}_{1:k} = [\mathbf{y}_1, \mathbf{y}_2, \dots, \mathbf{y}_k]$. Finally, the performances of the calibrated models are evaluated to select the model which is the best suited for its intended use.

The experimental calibration process is usually treated either from a frequentist or a Bayesian point of view. In Bayesian estimation, the unknown parameters are treated as random variables with a certain prior distribution $p(\boldsymbol{\theta})$, which represents the prior belief before looking at the data (Dahlin 2016). Then, all the information available in the data is summarized in the likelihood function $p(\mathbf{y}_{1:k}|\boldsymbol{\theta})$. The prior belief and the data information are combined in the Bayes' theorem to compute the posterior distribution:

$$p(\boldsymbol{\theta}|\mathbf{y}_{1:k}) = \frac{p(\mathbf{y}_{1:k}|\boldsymbol{\theta})p(\boldsymbol{\theta})}{p(\mathbf{y}_{1:k})} \propto p(\mathbf{y}_{1:k}|\boldsymbol{\theta})p(\boldsymbol{\theta}) \tag{1}$$

where $p(\mathbf{y}_{1:k})$ is a normalization constant independent of the parameters.

The posterior distribution $p(\boldsymbol{\theta}|\mathbf{y}_{1:k})$ contains all the statistical information about $\boldsymbol{\theta}$; the most probable value of the posterior distribution gives the *maximum a posteriori* (MAP) estimate:

$$\boldsymbol{\theta}_{\mathbf{MAP}} = \underset{\boldsymbol{\theta}}{\operatorname{argmax}}\left(p(\mathbf{y}_{1:k}|\boldsymbol{\theta})p(\boldsymbol{\theta})\right) \tag{2}$$

If only the information in the data is considered, maximizing the likelihood function $p(\mathbf{y}_{1:k}|\boldsymbol{\theta})$ gives the *maximum likelihood* (ML) estimate:

$$\boldsymbol{\theta}_{\mathbf{ML}} = \underset{\boldsymbol{\theta}}{\operatorname{argmax}}\left(p(\mathbf{y}_{1:k}|\boldsymbol{\theta})\right) \tag{3}$$

The ML estimate can be seen as a MAP estimate with uniform prior distribution, $p(\boldsymbol{\theta}) \propto 1$ (Sarkka 2013).

Two philosophies exist for computing these estimates and their uncertainties. The *frequentist* approach relies on the fact that, as the number of observations increases, the influence of the prior distribution becomes negligible compared to the likelihood and the posterior distribution can be approximated by a Gaussian distribution (Gelman, Carlin, et al. 2014). ML estimation is popular because it requires only point estimates of the posterior modes and their corresponding uncertainties are determined by asymptotic properties. ML estimates are usually found by optimization routines as in the CTSM-R package (CTSM-R Development Team 2015). This strategy has been proven to be efficient at numerous cases (Naveros et al. 2014, Himpe & Janssens 2015, Nespoli et al. 2015,

Andersen et al. 2014, Bacher & Madsen 2011, Váňa et al. 2013). However, the asymptotic theory does not hold for small number of observations, which is often the case in real experiment. From a Bayesian point of view, all the statistical information is summarized in the posterior distribution, thus no assumption is made.

Bayesian and frequentist methods are compared in the three phases of the calibration process in order to illustrate the differences and to take advantage of both methods.

### 3.1. Model selection

The choice of an appropriate model structure is the most crucial part in experimental calibration according to Ljung (2002). The model structure should be representative of the real physical system and also in agreement with the measured data. Based on a-priori knowledge of the experiment, a set of models of increasing complexity is defined which allows a forward selection between nested models, i.e. the simplest model can be recovered by putting extra parameters to zero (Bacher & Madsen 2011). In this paper, the distinction between model selection and model comparison is done because the model comparison requires to discuss about calibration first; therefore it is presented later. To illustrate the Bayesian experimental calibration process only the two most promising model candidates are presented. These two models, illustrated in Figure 2, have been selected from a larger set of models which is not presented here due to lack of space. The smallest thermal network in Figure 2, the 3 states model, $\mathcal{M}_3$, is obtained by not taking into account the surrounded dotted part whereas the 4 states model, $\mathcal{M}_4$, is obtained by adding this part to $\mathcal{M}_3$, such that, $\mathcal{M}_3 \subset \mathcal{M}_4$.

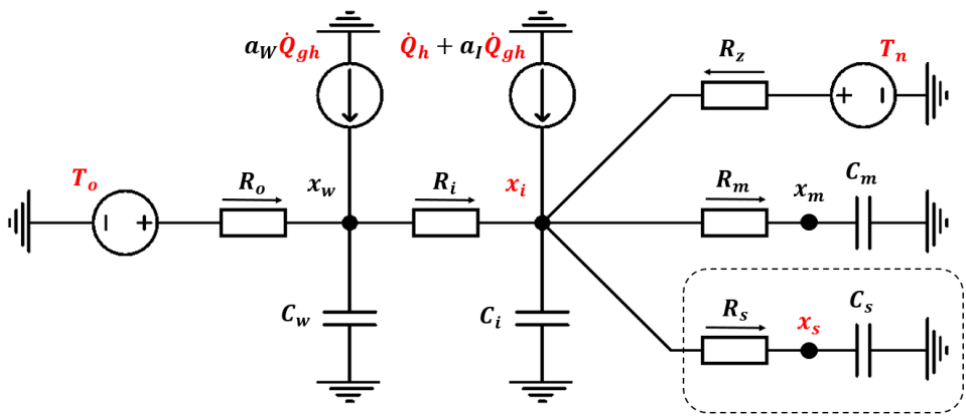

*Figure 2: 3 states model, $\mathcal{M}_3$ and 4 states model, $\mathcal{M}_4$*

In **Error! Reference source not found.** the nodes represent:

- $x_w$: the temperature of the building envelope (°C),
- $x_i$: the indoor air temperature (°C),
- $x_m$: the temperature of the medium (internal walls, furniture, etc) (°C),
- $x_s$: the temperature of the sensor (°C).

When the node $x_s$ is used, it is selected as the model output and is equivalent to the south zone air temperature measured $T_s$ (°C), otherwise $x_i$ is the model output.

The different heat transfers are characterized by the thermal resistances:

- $R_o$: between the outside and the middle of the building envelope (K/W),
- $R_i$: between half the building envelope and the south zone (K/W),
- $R_m$: between the south zone and the medium (K/W),
- $R_z$: between the air in the north zone and the south zone (K/W),
- $R_s$: between the south zone and the sensor (K/W).

The accumulation of energy is modeled by the thermal capacities:

- $C_w$: for the building envelope (J/K),
- $C_i$: for the air in the south zone (J/K),
- $C_m$: for the medium (internal walls, furniture, etc) (J/K),
- $C_s$: for the sensor.

The parameters $a_W$ and $a_I$ are respectively the effective area through which the solar radiation enters the building envelope and the effective window area of the building.

State-space models can be easily deduced from thermal networks by considering heat balance in each temperature nodes (Ghiaus 2013):

$$\dot{\mathbf{x}} = \mathbf{A}(\boldsymbol{\theta})\mathbf{x} + \mathbf{B}(\boldsymbol{\theta})\mathbf{u}$$
$$\mathbf{y}_k = \mathbf{C}\mathbf{x}_k$$
(4)

where $\mathbf{A}$ is the state matrix, $\mathbf{B}$ the input matrix, $\mathbf{C}$ the output matrix $\boldsymbol{\theta}$ the parameter vector, $\mathbf{u}$ the input vector and $\mathbf{y}_k$ is an output vector which can be measured at discrete time instant $t_k$ (e.g. $\mathbf{x}_k = \mathbf{x}(t_k)$).

Both models from Figure 2 share the same input vector

$$\mathbf{u} = \begin{bmatrix} T_o & T_z & a_W \dot{Q}_{gh} & a_I \dot{Q}_{gh} + \dot{Q}_h \end{bmatrix}^\mathrm{T}$$
(5)

The data of the twin houses experiment are provided with a constant sampling time $\Delta_t$; therefore, the linear time invariant continuous model (4) is discretized and additive noise terms are introduced to describe the deviation between the discrete system and the true variation of the state

$$\mathbf{x}_{k+1} = \mathbf{A_d}(\boldsymbol{\theta})\mathbf{x}_k + \mathbf{B_{d0}}(\boldsymbol{\theta})(\mathbf{u}_k + \boldsymbol{\alpha}\Delta_t) - \mathbf{B_{d1}}(\boldsymbol{\theta})\boldsymbol{\alpha} + \mathbf{w}_k$$

$$\mathbf{y}_k = \mathbf{C}\mathbf{x}_k + \mathbf{v}_k \tag{6}$$

where $\mathbf{w}_k$ and $\mathbf{v}_k$ are white noise processes with respective covariance $\boldsymbol{\Sigma_w}(\boldsymbol{\theta})$ and $\boldsymbol{\Sigma_v}(\boldsymbol{\theta})$, and

$$\mathbf{A_d}(\boldsymbol{\theta}) = e^{\mathbf{A}\Delta_t} \tag{6.a}$$

$$\mathbf{B_{d0}}(\boldsymbol{\theta}) = \mathbf{A}^{-1}(\mathbf{A_d} - \mathbf{I})\mathbf{B} \tag{6.b}$$

$$\mathbf{B_{d1}}(\boldsymbol{\theta}) = \mathbf{A}^{-1}(-\mathbf{A}^{-1}(\mathbf{A_d} - \mathbf{I}) + \mathbf{A_d}\Delta_t)\mathbf{B} \tag{6.c}$$

If the input is assumed constant in the time interval $\Delta_t$ (zero order hold), then $\boldsymbol{\alpha} = \mathbf{0}$, whereas if the input is assumed to vary linearly (first order hold), then (Kristensen & Madsen 2003)

$$\boldsymbol{\alpha} = \frac{\mathbf{u}_{k+1} - \mathbf{u}_k}{\Delta_t} \tag{7}$$

In order to make the notation lighter, the parameter dependence in $\mathbf{A_d}, \mathbf{B_{d0}}, \mathbf{B_{d1}}, \boldsymbol{\Sigma_w}$ and $\boldsymbol{\Sigma_v}$ is omitted such that $\mathbf{A_d} = \mathbf{A_d}(\boldsymbol{\theta})$.

Some parameters of the state-space model (6) may not be known and, then, they have to be estimated based on measured data.

### 3.2. Bayesian calibration with Markov Chain Monte Carlo

*Markov chain Monte Carlo* (MCMC) is a general method for constructing posterior distributions. The main idea is to simulate a Markov chain which has been constructed such that it has the posterior distribution as its stationary distribution (Sarkka 2013). The *Metropolis-Hastings* (MH) algorithm is the most common type of MCMC method due to its simplicity. MH is an iterative scheme, where a new candidate $\boldsymbol{\theta}^*$ is suggested from a proposed distribution $q(\boldsymbol{\theta}^*|\boldsymbol{\theta}^{i-1})$ given the previous one $\boldsymbol{\theta}^{i-1}$. The candidate is then accepted or rejected according to some acceptance probability

$$\alpha_i = \min\left\{1, \frac{p(\mathbf{y}_{1:k}|\boldsymbol{\theta}^*)p(\boldsymbol{\theta}^*)}{p(\mathbf{y}_{1:k}|\boldsymbol{\theta}^{i-1})p(\boldsymbol{\theta}^{i-1})} \frac{q(\boldsymbol{\theta}^{i-1}|\boldsymbol{\theta}^*)}{q(\boldsymbol{\theta}^*|\boldsymbol{\theta}^{i-1})}\right\} \tag{8}$$

where the ratio $q(\boldsymbol{\theta}^{i-1}|\boldsymbol{\theta}^*)/q(\boldsymbol{\theta}^*|\boldsymbol{\theta}^{i-1})$ corrects the asymmetry in the proposed distribution.

If the candidate $\boldsymbol{\theta}^*$ increases significantly the posterior probability, the candidate is always accepted. However, a candidate which decrease the posterior probability can still be accepted as opposed to optimization algorithms; it allows the MH algorithm to explore regions of high posterior probability. Hence, by its stochastic nature, the MH algorithm may escape from local extrema which is a problem for many optimization algorithms used for ML estimation (Dahlin 2016).

The performance of the MH algorithm is highly dependent on the choice of the proposed distribution. A commonly used choice is the Gaussian random walk,

$$q(\boldsymbol{\theta}^*|\boldsymbol{\theta}^{i-1}) = \mathcal{N}(\boldsymbol{\theta}^*|\boldsymbol{\theta}^{i-1}, \boldsymbol{\Sigma}_{\boldsymbol{\theta}}^{i-1}) \tag{9}$$

where $\mathcal{N}(\boldsymbol{\theta}^*|\boldsymbol{\theta}^{i-1}, \boldsymbol{\Sigma}_{\boldsymbol{\theta}}^{i-1})$ is a Gaussian probability density function of a random variable $\boldsymbol{\theta}^* \in \mathbb{R}^{N_p}$ with mean $\boldsymbol{\theta}^{i-1} \in \mathbb{R}^{N_p}$ and covariance $\boldsymbol{\Sigma}_{\boldsymbol{\theta}}^{i-1} \in \mathbb{R}^{N_p \times N_p}$,

$$\mathcal{N}(\boldsymbol{\theta}^*|\boldsymbol{\theta}^{i-1}, \boldsymbol{\Sigma}_{\boldsymbol{\theta}}^{i-1}) = \frac{1}{(2\pi)^{n/2}|\boldsymbol{\Sigma}_{\boldsymbol{\theta}}^{i-1}|^{1/2}} \exp\left(-\frac{1}{2}(\boldsymbol{\theta}^* - \boldsymbol{\theta}^{i-1})^{\mathrm{T}}(\boldsymbol{\Sigma}_{\boldsymbol{\theta}}^{i-1})^{-1}(\boldsymbol{\theta}^* - \boldsymbol{\theta}^{i-1})\right) \tag{10}$$

Finding a suitable covariance matrix $\boldsymbol{\Sigma}$ is a hard task which involves many trials and becomes unrealistic for high-dimensional problems (Sarkka 2013). The Markov chain should converge to the stationary distribution in a reasonable time (burn-in phase) and the Markov chain should not be highly autocorrelated, such that the number of iteration for exploring the stationary distribution is minimized.

The performance and the tuning of the MH algorithm can be respectively improved and simplified by using the gradient and Hessian of the posterior distribution in order to construct a better proposed distribution (Dahlin 2016). The next section shows a robust and accurate method for computing the likelihood, gradient and Hessian for linear Gaussian state-space models (6).

### 3.2.1. Construction of the posterior and proposal distribution

The construction of the posterior distribution (1) requires the evaluation of the likelihood $p(\mathbf{y}_{1:k}|\boldsymbol{\theta})$ and the prior distribution $p(\boldsymbol{\theta})$. The challenging part is the evaluation of the likelihood because the prior distribution is usually chosen such that it is easy to evaluate (Sarkka 2013). For a state-space model, the likelihood can be computed by using the prediction error decomposition

$$p(\mathbf{y}_{1:k}|\boldsymbol{\theta}) = p(\mathbf{y}_1|\boldsymbol{\theta}) \prod_{k=2}^{N} p(\mathbf{y}_k|\mathbf{y}_{1:k-1}, \boldsymbol{\theta}) \tag{11}$$

where the predictive likelihood can be computed recursively by

$$p(\mathbf{y}_k|\mathbf{y}_{1:k-1}, \boldsymbol{\theta}) = \int p(\mathbf{y}_k|\mathbf{x}_k, \boldsymbol{\theta}) \, p(\mathbf{x}_k|\mathbf{y}_{1:k-1}, \boldsymbol{\theta}) \, d\mathbf{x}_k \tag{11.a}$$

with $p(\mathbf{y}_k|\mathbf{x}_k, \boldsymbol{\theta})$ and $p(\mathbf{x}_k|\mathbf{y}_{1:k-1}, \boldsymbol{\theta})$ representing respectively the measurement model and the predictive distribution of the state.

To avoid computational inaccuracy and instability, the logarithm of the unnormalized posterior distribution (right-hand side of (1)), named log-posterior, is computed instead.

Since not all the states are not observed, the parameter estimation problem also requires to solve the state estimation problem (Dahlin 2016). For linear Gaussian state-space model, the integral in (11.a) can be computed in closed form by the Kalman filter. The gradient and an approximation of the Hessian are obtained by differentiation of the Kalman filter equations with respect to the unknown parameters, referred to as the sensitivity equations. Therefore, computation of $N_p$ sensitivity equations is required in parallel to the Kalman filter, where $N_p$ is the number of unknown parameters. However, this strategy is numerically unstable due to rounding errors, i.e., the state covariance matrix $\mathbf{P}$ may cease to be symmetric and positive definite, which leads to the failure of the computational process. This problem can be solved by using a robust square root implementation (Kulikova & Tsyganova 2016, Tsyganova & Kulikova 2012), where only the square root factor $\mathbf{P}^{1/2}$ is propagated instead of the full state covariance matrix. The upper triangular matrix $\mathbf{P}^{1/2}$ is obtained by *Cholesky decomposition*, such that $\mathbf{P} = \mathbf{P}^{T/2}\mathbf{P}^{1/2}$.

The recursion starts by updating the mean value of the prior state $\mathbf{x}_{k|k-1}$ and prior square root factor $\mathbf{P}^{1/2}_{k|k-1}$ with the current measurements

$$\mathbf{x}_{k|k} = \mathbf{x}_{k|k-1} + \overline{\mathbf{K}}_k \overline{\mathbf{e}}_k \tag{12}$$

with the standardized residuals

$$\overline{\mathbf{e}}_k = \mathbf{S}_k^{-1/2}(\mathbf{y}_k - \mathbf{C}_\mathbf{d}\mathbf{x}_{k|k-1}) \tag{13}$$

The normalized Kalman gain $\overline{\mathbf{K}}_k$ and the square root factor of the residual covariance, $\mathbf{S}_k^{1/2}$, are directly read from the post-array, in the right-hand side of equation (14)

$$\mathbf{Q}_k^{\mathrm{MU}} \underbrace{\begin{bmatrix} \boldsymbol{\Sigma}_\mathbf{v}^{1/2} & 0 \\ \mathbf{P}^{1/2}_{k|k-1}\mathbf{C}_\mathbf{d}^T & \mathbf{P}^{1/2}_{k|k-1} \end{bmatrix}}_{\text{Pre-array}} = \underbrace{\begin{bmatrix} \mathbf{S}_k^{1/2} & \overline{\mathbf{K}}_k^T \\ 0 & \mathbf{P}^{1/2}_{k|k} \end{bmatrix}}_{\text{Post-array}} \tag{14}$$

where $\mathbf{Q}_k^{\mathrm{MU}}$ is an orthogonal rotation matrix obtained by *QR decomposition* of the pre-array, such that the post-array is upper triangular; this notation is used throughout the paper.

The partial derivatives of the quantities in the post-array (14) are computed with

$$\frac{\partial S_k^{1/2}}{\partial \theta_i} = \left(\left(\mathcal{L}_i^{\ddagger}\right)^T + \mathcal{D}_i^{\ddagger} + \mathcal{U}_i^{\ddagger}\right) S_k^{1/2} \tag{15}$$

$$\frac{\partial \bar{K}_k}{\partial \theta_i} = Y_i^T + \left(\left(\mathcal{L}_i^{\ddagger}\right)^T - \mathcal{L}_i^{\ddagger}\right) \bar{K}_k^T + \left(V_i S_k^{-1/2}\right)^T P_{k|k}^{1/2} \tag{16}$$

$$\frac{\partial P_{k|k}^{1/2}}{\partial \theta_i} = \left(\left(\mathcal{L}_i^{\dagger}\right)^T + \mathcal{D}_i^{\dagger} + \mathcal{U}_i^{\dagger}\right) P_{k|k}^{1/2} \tag{17}$$

Where $\mathcal{L}^{\dagger}, \mathcal{D}^{\dagger}, \mathcal{U}^{\dagger}, \mathcal{L}^{\ddagger}, \mathcal{D}^{\ddagger}$ and $\mathcal{U}^{\ddagger}$ are obtained by first multiplying the orthogonal rotation matrix $Q_k^{MU}$ to the partial derivative of the pre-array in (14)

$$Q_k^{MU} \begin{bmatrix} \frac{\partial \Sigma_v^{1/2}}{\partial \theta_i} & 0 \\ \frac{\partial \left(P_{k|k-1}^{1/2} C_d^T\right)}{\partial \theta_i} & \frac{\partial P_{k|k-1}^{1/2}}{\partial \theta_i} \end{bmatrix} = \begin{bmatrix} X_i & Y_i \\ V_i & W_i \end{bmatrix} \tag{18}$$

and then the post-array (18) is multiplied by the inverse of the pre-array (14)

$$G = \begin{bmatrix} X_i & Y_i \\ V_i & W_i \end{bmatrix} \begin{bmatrix} S_k^{1/2} & \bar{K}_k^T \\ 0 & P_{k|k}^{1/2} \end{bmatrix}^{-1} \tag{19}$$

The matrices $\mathcal{L}^{\dagger}, \mathcal{D}^{\dagger}$ and $\mathcal{U}^{\dagger}$ are respectively the lower triangular, diagonal, and upper triangular parts of the submatrix G, from row and column $N_y + 1$ to row and column $N_y + N_x$. Matrices $\mathcal{L}^{\ddagger}, \mathcal{D}^{\ddagger}$ and $\mathcal{U}^{\ddagger}$ are respectively the lower triangular, diagonal, and upper triangular parts of the submatrix G, from the first row and column to row and column $N_y$ (Tsyganova & Kulikova 2012).

The partial derivative of the prior state mean update is computed by

$$\frac{\partial x_{k|k}}{\partial \theta_i} = \frac{\partial x_{k|k-1}}{\partial \theta_i} + \frac{\partial \bar{K}_k}{\partial \theta_i} \bar{e}_k + \bar{K}_k \frac{\partial \bar{e}_k}{\partial \theta_i} \tag{20}$$

with

$$\frac{\partial \bar{e}_k}{\partial \theta_i} = -S_k^{-1/2} \left(\frac{\partial S_k^{1/2}}{\partial \theta_i} \bar{e}_k + \frac{\partial C_d}{\partial \theta_i} x_{k|k-1} + C_d \frac{\partial x_{k|k-1}}{\partial \theta_i}\right) \tag{21}$$

The posterior state mean and square root factor are propagated forward in time by the state equation similar to (6)

$$x_{k+1|k} = A_d x_{k|k} + B_{d0}(u_k + \alpha \Delta_t) - B_{d1} \alpha \tag{22}$$

and by

$$\mathbf{Q}_k^{\text{TU}} \begin{bmatrix} \mathbf{P}_{k|k}^{1/2} \mathbf{A}_\mathbf{d}^\text{T} \\ \mathbf{\Sigma}_\mathbf{w}^{1/2} \end{bmatrix} = \begin{bmatrix} \mathbf{P}_{k+1|k}^{1/2} \\ 0 \end{bmatrix} \tag{23}$$

The partial derivative of the state equation (22) is

$$\frac{\partial \mathbf{x}_{k+1|k}}{\partial \theta_i} = \frac{\partial \mathbf{A}_\mathbf{d}}{\partial \theta_i} \mathbf{x}_{k|k} + \mathbf{A}_\mathbf{d} \frac{\partial \mathbf{x}_{k|k}}{\partial \theta_i} + \frac{\partial \mathbf{B}_{\mathbf{d}0}}{\partial \theta_i} (\mathbf{u}_{k-1} + \boldsymbol{\alpha}\Delta_t) - \frac{\partial \mathbf{B}_{\mathbf{d}1}}{\partial \theta_i} \boldsymbol{\alpha} \tag{24}$$

where the partial derivatives of the state and input discrete matrices with respect to the continuous parameters are computed by (Mbalawata et al. 2013)

$$\underbrace{\begin{bmatrix} \mathbf{A}_\mathbf{d} & 0 \\ \frac{\partial \mathbf{A}_\mathbf{d}}{\partial \theta_i} & \mathbf{A}_\mathbf{d} \end{bmatrix}}_{\mathbf{A}_{\text{IM}}} = \exp\left( \underbrace{\begin{bmatrix} \mathbf{A} & 0 \\ \frac{\partial \mathbf{A}}{\partial \theta_i} & \mathbf{A} \end{bmatrix}}_{\mathbf{A}_\mathbf{M}} \Delta t \right) \tag{24.a}$$

$$\begin{bmatrix} \mathbf{B}_{\mathbf{d}0} \\ \frac{\partial \mathbf{B}_{\mathbf{d}0}}{\partial \theta_i} \end{bmatrix} = \mathbf{A}_\mathbf{M}^{-1} (\mathbf{A}_{\text{IM}} - \mathbf{I}) \begin{bmatrix} \mathbf{B} \\ \frac{\partial \mathbf{B}}{\partial \theta_i} \end{bmatrix} \tag{24.b}$$

$$\begin{bmatrix} \mathbf{B}_{\mathbf{d}1} \\ \frac{\partial \mathbf{B}_{\mathbf{d}1}}{\partial \theta_i} \end{bmatrix} = \mathbf{A}_\mathbf{M}^{-1} (-\mathbf{A}_\mathbf{M}^{-1}(\mathbf{A}_{\text{IM}} - \mathbf{I}) + \mathbf{A}_{\text{IM}} \Delta_t) \begin{bmatrix} \mathbf{B} \\ \frac{\partial \mathbf{B}}{\partial \theta_i} \end{bmatrix} \tag{24.c}$$

The partial derivative of the square root factor in the post-array (23)

$$\frac{\partial \mathbf{P}_{k+1|k}^{1/2}}{\partial \theta_i} = (\mathcal{L}_i^\text{T} + \mathcal{D}_i + \mathcal{U}_i) \mathbf{P}_{k+1|k}^{1/2} \tag{25}$$

requires to first multiply the orthogonal rotation matrix $\mathbf{Q}_k^{\text{TU}}$ by the partial derivative of the pre-array (23)

$$\mathbf{Q}_k^{\text{TU}} \begin{bmatrix} \frac{\partial (\mathbf{P}_{k|k}^{1/2} \mathbf{A}_\mathbf{d}^\text{T})}{\partial \theta_i} \\ \frac{\partial \mathbf{\Sigma}_\mathbf{w}^{1/2}}{\partial \theta_i} \end{bmatrix} = \begin{bmatrix} \mathbf{A}_i \\ \mathbf{C}_i \end{bmatrix} \tag{26}$$

where the matrices $\mathcal{L}$, $\mathcal{D}$ and $\mathcal{U}$ are respectively the lower triangular, diagonal, and upper triangular parts of the matrix product $\mathbf{A}_i \mathbf{P}_{k|k}^{-\text{T}/2}$.

The log-likelihood is recursively computed with

$$\ln p(\mathbf{y}_{1:k}|\boldsymbol{\theta}) = \frac{1}{2} \sum_{k=1}^{N} -\frac{N_y}{2} \ln(2\pi) - \ln\left(\det(\mathbf{S}_k^{\text{T}/2} \mathbf{S}_k^{1/2})\right) - \mathbf{e}_k^\text{T} \bar{\mathbf{e}}_k \tag{27}$$

where the standardized innovations $\bar{\mathbf{e}}_k$ and the square-root of the innovation covariance matrix $\mathbf{S}_k^{1/2}$ are computed respectively by (13) and (14).

The gradient and the Hessian approximation of (27) are respectively obtained with

$$\frac{\partial \ln p(\mathbf{y}_{1:k}|\boldsymbol{\theta})}{\partial \theta_i} = -\sum_{k=1}^{N} \text{tr}\left(\mathbf{S}_k^{-1/2} \frac{\partial \mathbf{S}_k^{1/2}}{\partial \theta_i}\right) + \frac{\partial \bar{\mathbf{e}}_k^T}{\partial \theta_i} \bar{\mathbf{e}}_k \tag{28}$$

$$-\frac{\partial^2 \ln p(\mathbf{y}_{1:k}|\boldsymbol{\theta})}{\partial \theta_i \partial \theta_j} \approx \sum_{k=1}^{N} \text{tr}\left(\frac{\partial \bar{\mathbf{e}}_k}{\partial \theta_i} \frac{\partial \bar{\mathbf{e}}_k^T}{\partial \theta_j}\right) + \text{tr}\left(\mathbf{S}_k^{-1/2} \frac{\partial \mathbf{S}_k^{1/2}}{\partial \theta_i} \mathbf{S}_k^{-1/2} \frac{\partial \mathbf{S}_k^{1/2}}{\partial \theta_j}\right) \tag{29}$$

It has been shown in this section that the parameter estimation problem is also a state estimation problem. For a linear Gaussian state-space model, the log-likelihood with its gradient and Hessian approximation can be computed by a square root version of the Kalman filter. This numerically stable strategy requires only to run the square root Kalman filer and $N_p$ sensitivity equations forward in time.

The gradient and the Hessian of the log-posterior distribution are easily computed with

$$\mathbf{g}(\boldsymbol{\theta}) = \frac{\partial \ln p(\boldsymbol{\theta}|\mathbf{y}_{1:k})}{\partial \boldsymbol{\theta}} + \frac{\partial \ln p(\boldsymbol{\theta})}{\partial \boldsymbol{\theta}} \tag{30}$$

$$\widehat{\mathbf{H}}(\boldsymbol{\theta}) = -\frac{\partial^2 \ln p(\boldsymbol{\theta}|\mathbf{y}_{1:k})}{\partial^2 \boldsymbol{\theta}} - \frac{\partial^2 \ln p(\boldsymbol{\theta})}{\partial^2 \boldsymbol{\theta}} \tag{31}$$

and are used in the MH algorithm to construct an efficient proposal distribution (Dahlin 2016)

$$q(\boldsymbol{\theta}^*|\boldsymbol{\theta}^{i-1}) = \mathcal{N}\left(\boldsymbol{\theta}^*\middle|\boldsymbol{\theta}^{i-1} + \frac{1}{2}\boldsymbol{\Sigma}\widehat{\mathbf{H}}^{-1}(\boldsymbol{\theta}^{i-1})\mathbf{g}(\boldsymbol{\theta}^{i-1}), \boldsymbol{\Sigma}\widehat{\mathbf{H}}^{-1}(\boldsymbol{\theta}^{i-1})\right) \tag{32}$$

where $\boldsymbol{\Sigma}$ is a diagonal matrix which control the step length of the proposal distribution.

Using the geometric information of the posterior distribution has the advantage of steering the Markov chain towards areas of high posterior probability (Nemeth 2014), which reduces the burn-in phase since the Markov chain takes larger steps when the Markov chain is far from the posterior mode and smaller steps as it gets closer. This approach allows to save user time and also computational time. Firstly, because the covariance matrix is given by the inverse of the Hessian approximation, thus only the step length has to be tuned. Secondly, because it increases the mixing of the Markov chain, so the MH algorithm needs less iterations (Dahlin 2016, Nemeth 2014).

The proposal distribution (32) is based on Newton-type optimization; consequently, the suggested candidates $\boldsymbol{\theta}^*$ are unconstrained and can violate the physical meaning of the system. A simple solution would be to reject candidates outside specified bounds but it could increase the autocorrelation of the Markov chain if too many candidates are rejected. A better solution is to reparametrize the model (Dahlin 2016).

### 3.2.2. Reparametrization of the model

The idea is to use non-linear functions to transform a constrained problem into an unconstrained one. In this way, the proposal distribution cannot suggest candidates which are outside the bounds. The constrained parameters $\boldsymbol{\theta}$ are transformed to unconstrained parameters $\boldsymbol{\eta}$ by a one-to-one invertible functions $\boldsymbol{\eta} = \mathbf{f}(\boldsymbol{\theta})$. Two parametrizations are used: 1) the log transform

$$\eta_i = \ln(\theta_i) \tag{33}$$

constraints $\theta_i$ between the open interval $]0, +\infty[$ and 2) the following transformation (Team Stan Development 2015)

$$\eta_j = \text{logit}\left(\frac{\theta_j - \theta_j^{\min}}{\theta_j^{\max} - \theta_j^{\min}}\right) \tag{34}$$

constraints $\theta_j$ between the open interval $]\theta_j^{\min}, \theta_j^{\max}[$, where the logit function is

$$\text{logit}(z) = \ln\left(\frac{z}{1-z}\right) \tag{34.a}$$

In the acceptance probability (8), the unnormalized posterior distribution is computed in the constrained space whereas the proposal distribution is evaluated in the unconstrained one. To homogenize the acceptance probability, the log-posterior distribution is transformed in the unconstrained space by using the Jacobian adjustment (Gelman, Carlin, et al. 2014)

$$\ln p(\boldsymbol{\eta}|\mathbf{y}_{1:k}) = \ln p(\boldsymbol{\theta}|\mathbf{y}_{1:k}) + \ln|\det(\mathbf{J})| \tag{35}$$

where $|\det(\mathbf{J})|$ is the absolute value of the determinant of the Jacobian matrix $\mathbf{J}$; $|\det(\mathbf{J})|$ adjusts for the distortion caused by the non-linear transformation; $\mathbf{J}$ is the Jacobian matrix of the inverse transform $\boldsymbol{\theta} = \mathbf{f}^{-1}(\boldsymbol{\eta})$, such that

$$J_{ij} = \frac{\partial \theta_i}{\partial \eta_j} \tag{35.a}$$

The Jacobian matrix is triangular if each transformed parameter only depends on a single untransformed parameter, which simplifies the determinant computation to the product of the diagonal elements.

The gradient (30) and the Hessian (31) are with respect to $\boldsymbol{\theta}$, so they need to be multiplied by the Jacobian $\mathbf{J}$ in order to be with respect to $\boldsymbol{\eta}$ (chain rule). The partial derivative of the new term in (35) needs also to be taken into account. The gradient and the Hessian in the unconstrained space are obtained by

$$\mathbf{g}(\boldsymbol{\eta}) = \mathbf{J}^T \mathbf{g}(\boldsymbol{\theta}) + \frac{\partial \ln|\det(\mathbf{J})|}{\partial \boldsymbol{\eta}} \tag{36}$$

$$\widehat{\mathbf{H}}(\boldsymbol{\eta}) = \mathbf{J}^T \widehat{\mathbf{H}}(\boldsymbol{\theta}) \mathbf{J} + \frac{\partial^2 \ln|\det(\mathbf{J})|}{\partial^2 \boldsymbol{\eta}} \tag{37}$$

A problem arises with this type of reparametrization when $\theta_i$ gets close to a bound: its corresponding Jacobian term goes towards zero, so $\mathbf{g}(\boldsymbol{\eta})$ and $\widehat{\mathbf{H}}(\boldsymbol{\eta})$ are unreliable. Moreover, the Hessian estimate $\widehat{\mathbf{H}}(\boldsymbol{\eta})$ could become ill-conditioned which prevents from its inversion. This issue is solved in optimization by adding a penalty function (Kristensen & Madsen 2003) which increases the gradient near the bounds, but this strategy is not suitable here because the proposal distribution (32) has a stochastic part, which means that candidates can still be projected towards the bounds. Instead, a prior distribution $p(\boldsymbol{\theta})$ is used to assign a low probability near the bounds. The details of the MH algorithm with gradient and Hessian information for constrained parameters are given in Algorithm 1.

---

**Algorithm 1: Second order Metropolis-Hastings** (Dahlin 2016)

**Inputs:** $N$ (number of iterations), $\boldsymbol{\theta}^0$ (initial parameters), $\boldsymbol{\Sigma}$ (step length)

**Output:** $\boldsymbol{\eta}^{1:N}$ (samples from the posterior distribution)

---

1. Transformation to the unconstrained parameter space $\boldsymbol{\eta}^0 = \mathbf{f}(\boldsymbol{\theta}^0)$ with (33) and (34)
2. Compute: $\ln p(\boldsymbol{\eta}^0|\mathbf{y}_{1:k})$, $\mathbf{g}(\boldsymbol{\eta}^0)$ and $\widehat{\mathbf{H}}(\boldsymbol{\eta}^0)$ with (35), (36) and (37)
3. $for\ i = 1\ to\ N$
4.     Suggest a new candidate $\boldsymbol{\eta}^* \sim \mathcal{N}\left(\boldsymbol{\eta}^{i-1} + \frac{1}{2}\boldsymbol{\Sigma}\widehat{\mathbf{H}}^{-1}(\boldsymbol{\eta}^{i-1})\mathbf{g}(\boldsymbol{\eta}^{i-1}), \boldsymbol{\Sigma}\widehat{\mathbf{H}}^{-1}(\boldsymbol{\eta}^{i-1})\right)$
5.     Compute: $\ln p(\boldsymbol{\eta}^*|\mathbf{y}_{1:k})$, $\mathbf{g}(\boldsymbol{\eta}^*)$ and $\widehat{\mathbf{H}}(\boldsymbol{\eta}^*)$ with (35), (36) and (37)
6.     Compute the acceptance probability $\alpha_i$ with (8)
7.     Generate a uniform random variable $u \sim \mathcal{U}(0,1)$ and set
8.     $if\ u \leq \alpha_i$
9.         Accept the new candidate
   $\{\boldsymbol{\eta}^i, \ln p(\boldsymbol{\eta}^i|\mathbf{y}_{1:k}), \mathbf{g}(\boldsymbol{\eta}^i), \widehat{\mathbf{H}}(\boldsymbol{\eta}^i)\} \leftarrow \{\boldsymbol{\eta}^*, \ln p(\boldsymbol{\eta}^*|\mathbf{y}_{1:k}), \mathbf{g}(\boldsymbol{\eta}^*), \widehat{\mathbf{H}}(\boldsymbol{\eta}^*)\}$
10.     $else$
11.         Reject the new candidate
    $\{\boldsymbol{\eta}^i, \ln p(\boldsymbol{\eta}^i|\mathbf{y}_{1:k}), \mathbf{g}(\boldsymbol{\eta}^i), \widehat{\mathbf{H}}(\boldsymbol{\eta}^i)\} \leftarrow \{\boldsymbol{\eta}^{i-1}, \ln p(\boldsymbol{\eta}^{i-1}|\mathbf{y}_{1:k}), \mathbf{g}(\boldsymbol{\eta}^{i-1}), \widehat{\mathbf{H}}(\boldsymbol{\eta}^{i-1})\}$
12.     $end\ if$
13. $end\ for$

---

### 3.2.3. Choice of prior distribution

The knowledge of possible parameter values before anything has been observed is represented probabilistically by the prior distribution $p(\boldsymbol{\theta})$. A prior distribution, which is relevant with the experiment, the physical nature of the problem, or for other reasons, has to be specified by the user. Three categories of prior information are considered: non-informative, weakly informative and informative (see Gelman et al. 2014, for a complete discussion on the subject).

Non-informative prior distributions attempt to not affect the posterior distribution, such that only the information in the data are relevant; this is the idea behind the ML estimation. But, these flat or almost flat prior distributions put more probability mass outside the expected range of values than inside, which can have unforeseen effect on the posterior distribution (Dahlin 2016), especially for small data set. Moreover, non-informative prior distributions, such as $\mathcal{U}(-\infty, +\infty)$ may be improper (they do not integrate to one), thus they cannot be expressed as a probability density function. In some cases, proper posterior distribution can be obtained with an improper prior distribution, but the result must be interpreted with care (Gelman, Carlin, et al. 2014).

Weakly informative prior distributions provide sufficient information to keep the parameters in a reasonable range and unlike informative prior distributions, they are not likely to outweigh the likelihood. For cases where the data set is too short or not enough informative, weakly informative prior distribution contains enough information to regularize the posterior distribution and prevent from identifiability issues; the curvature around the expected solution is increased (Team Stan Development 2015).

For the parameters transformed with the logarithm (33), the prior distributions are $\theta_i \sim \mathcal{G}(a, b)$, where $\mathcal{G}(a, b)$ denotes a Gamma distribution with shape $a$ and expected value $b$. The hyper-parameters $a$ and $b$ are chosen such that the probability near zero is low and that the distribution covers the expected range of values (Figure 3). For the transformed parameter with lower and upper bounds, the prior distributions are $\theta_j \sim \mathcal{B}(2, 2, \theta_j^{\min}, \theta_j^{\max})$, where $\mathcal{B}(a, b, \theta_j^{\min}, \theta_j^{\max})$ is a Beta distribution with shape hyper-parameters $a$ and $b$, lower and upper bounds $\theta_j^{\min}$ and $\theta_j^{\max}$. The Beta distribution with $a = b = 2$ is symmetric and assigns low probabilities for values near the bounds (Figure 3).

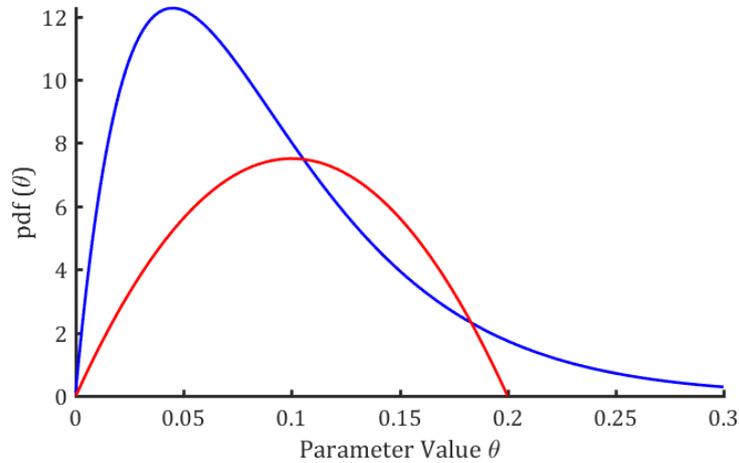

*Figure 3: Prior pdf, $\mathcal{G}(2, 0.03)$ in blue and $\mathcal{B}(2, 2, 10^{-4}, 2 \cdot 10^{-1})$ in red*

### 3.2.4. Tuning the algorithm

The choice of the prior distribution is an important decision which can strongly influence the posterior distribution (Dahlin 2016) but the exploration of the posterior distribution depends on the tuning of the algorithm. The use of the Hessian approximation reduces the tuning of the proposal distribution to the choice of the step length matrix $\boldsymbol{\Sigma}$

(32). Separate step lengths for each parameter can be used, but to simplify the tuning, when a single step length is used, such that $\boldsymbol{\Sigma} = \varepsilon \mathbf{I}_{Np}$. The step length affects directly the acceptance rate of the MH algorithm (the percentage of accepted candidates at stationarity). A too large $\boldsymbol{\Sigma}$ produces broad jumps which are more likely to be rejected, which increases the autocorrelation of the Markov chains and give a low acceptance rate. On the contrary, if the step length is too small, short jumps are likely to be accepted, which gives a higher acceptance rate. However, it limits the exploration of the posterior distribution to a small neighborhood, which also increases the autocorrelation. Consequently, the acceptance rate alone is not a correct indicator of the algorithm performance.

A better solution is to look at the mixing of the Markov chains at stationarity, which can be quantified by the integrated autocorrelation time (IACT) (Dahlin 2016)

$$\widehat{\text{IACT}}\left(\theta_j^{N_b:N}\right) = 1 + 2 \sum_{l=1}^{L} \hat{\rho}_l\left(\theta_j^{N_b:N}\right) \tag{38}$$

where $\rho_l$ denotes the autocorrelation coefficient at lag $l$ of $\theta_j^{N_b:N}$, and $\theta_j^{N_b:N}$ is the Markov chain of $\theta_j$ from the burn-in time $N_b$ to the last iteration $N$. The number of lags $L$ is determined as the first index for which $\left|\hat{\rho}_l\left(\theta_j^{N_b:N}\right)\right| < 2/\sqrt{N - N_b}$, when the autocorrelation coefficient becomes statistically insignificant.

The IACT represents the number of iterations between two uncorrelated samples; therefore, the step length should be chosen such that it minimizes the IACT. The number of iteration $N$ should be chosen sufficiently large, such that, once the burn-in phase has been removed, the number of samples left are sufficient to represent the posterior distribution. Tools for diagnosing convergence are discussed in the next section.

### 3.2.5. Convergence diagnosis

The procedure of Gelman et al. (2014) is used here and presented briefly. The procedure consists of simulating $M$ Markov chains of $N$ samples, where the starting points of the $M$ Markov chains are randomly sampled from the prior distributions. The first step consists of inspecting visually the trace plots of the different Markov chains to determine the burn-in time $N_b$ and to check if they converge to the same posterior distributions.

The $M$ Markov chains, with the burn-in phase removed, are split in two to give $2M$ chains of length $(N - N_b)/2$; then the variations between and within the $2M$ chains are compared (Gelman, Carlin, et al. 2014). The stationarity implies that the first and the second half of each sequence come from the same distribution. A good mixing requires that the variance inside chains should be closed to the variance between chains; this is quantified by the potential scale reduction $\hat{R}$ (see Gelman et al. 2014 for computational details). The number of iterations $N$ should be increased until $\hat{R}$ is near one or at least $\hat{R} < 1.1$. The mixing of the Markov chains can also be quantified by the effective sample size (ESS) which approximates the number of independent samples in the $2M$ sequences. Gelman et al. (2014) suggest that the ESS should be at least superior to $5 \times 2M$. If the aforementioned criteria are satisfied, the samples from the $2M$ sequences can be used to estimate the posterior distribution.

### 3.3. Model validation

After having calibrated different models, how to assess their reliability? The purpose of a model is to reproduce an input-output relationship; an intuitive way to start is by looking at what the model is not able to reproduce, the residuals $\bar{\mathbf{e}}$ (13) (Ljung 2002). A plot of the residuals and the data allows to understand which features are not properly described by the model and who might be responsible for. The noise terms in model (6) are assumed to be white noise sequences which implies that it should also be the case for the residuals. The white noise sequence is uncorrelated, normally distributed with a zero mean and is uniformly distributed on all frequencies (Madsen 2007). These properties are assessed by plotting the autocorrelation function (ACF) and the cumulated periodogram (CP) with their respective 95% confidence intervals. Furthermore, the residuals should be independent of the past inputs, which is tested by plotting the cross-correlation function (CCF).

The reliability of the model is also tested on a data set which has not been used for the calibration (validation data set). New values for the inputs are introduced in the model and the simulated output is compared to the measured one. If the identification data set is informative enough, i.e. the different dynamics of the system are observable in the data, the model should be representative of the system and therefore the simulation should be close to the measurement.

The model validation assesses the reliability of a model and gives insight on the model order selection and directions for improvement; but how to select the best model?

### 3.4. Model comparison

In section 3.1, the selection of a model structure based on insights of the experiment has been presented. This section discusses the agreement of calibrated models with measured data and how to compare different models in order to select the most appropriate one. The model fit to the data is summarized by the log-likelihood $\ln p(\mathbf{y}_{1:k}|\boldsymbol{\theta})$ (27) and the prior distribution is not relevant for assessing the accuracy of a model. The best model is not necessarily characterized by the highest log-likelihood value because as the complexity of a model increases, the number of degrees of freedom increases, the parameters adjust themselves to fit a particular realization of the noise (overfitting) (Ljung 2002). In order to adjust for overfitting, the *Akaike's Information Criterion* (AIC) and *Bayesian Information Criterion* (BIC) penalize the log-likelihood in function of the complexity of the model:

$$\text{AIC} = -2 \ln p(\mathbf{y}_{1:k}|\boldsymbol{\theta}_{\mathbf{ML}}) + 2N_p \quad (39)$$
$$\text{BIC} = -2 \ln p(\mathbf{y}_{1:k}|\boldsymbol{\theta}_{\mathbf{ML}}) + N_p \ln N_s \quad (40)$$

where $\boldsymbol{\theta}_{\mathbf{ML}}$ is the ML estimate, $N_p$ the number of parameters and $N_s$ the sample size.

The smallest AIC or BIC between different models indicates the most appropriate model.

For nested models, like $\mathcal{M}_3$ and $\mathcal{M}_4$, the *likelihood ratio test* (LRT) can be used (Bacher & Madsen 2011)

$$\text{LRT} = -2\left(\ln p(\mathbf{y}_{1:k}|\boldsymbol{\theta}_{\mathbf{ML}}^{\mathcal{M}_3}) - \ln p(\mathbf{y}_{1:k}|\boldsymbol{\theta}_{\mathbf{ML}}^{\mathcal{M}_4})\right) \quad (41)$$

with $\boldsymbol{\theta}_{\mathbf{ML}}^{\mathcal{M}_3}$ and $\boldsymbol{\theta}_{\mathbf{ML}}^{\mathcal{M}_4}$ the ML estimate of model $\mathcal{M}_3$ and $\mathcal{M}_4$.

As the number of samples $N_s$ goes to infinity, the LRT converges to $\chi^2$ distributed variable with $\left(N_p^{\mathcal{M}_4} - N_p^{\mathcal{M}_3}\right)$ degrees of freedom. Usually, a $p_{Value}$ of the LRT below 0.05, indicates that the improvement of the larger model $\mathcal{M}_4$ over $\mathcal{M}_3$ is significant and consequently the model $\mathcal{M}_4$ should be preferred.

These criteria are based on point estimate $\boldsymbol{\theta}_{\mathbf{ML}}$ and not on the posterior distribution $p(\boldsymbol{\theta}|\mathbf{y}_{1:k})$; a more Bayesian criterion is given by the *Watanabe-Akaike Information Criterio* (WAIC) (Gelman, Hwang, et al. 2014)

$$\text{WAIC} = -2 \sum_{k=1}^{N_s} \text{mean}(\ln p(\mathbf{y}_k|\boldsymbol{\theta})) - \text{var}(\ln p(\mathbf{y}_k|\boldsymbol{\theta})) \quad (42)$$

where $\ln p(\mathbf{y}_k|\boldsymbol{\theta}) \in \mathbb{R}^{(N-N_b) \times N_s}$ is the log-likelihood at time instant $k$ and $(N - N_b)$ is the number of sample used to approximate the posterior distribution $\ln p(\boldsymbol{\theta}|\mathbf{y}_{1:k})$. Differently of the AIC and BIC, the WAIC is penalized by the dispersion of the log-likelihood.

Evaluating these criteria on same data set used for the calibration introduced a bias in the model selection process and therefore it is advised to evaluate them with the validation data set instead; this point is illustrated in the next section for a real test case.

## 4. Application to the Twin houses experiment

### 4.1. Model comparison

The capabilities of the second order MH (Algorithm 1) are now tested on the twin houses experiment presented in section 2. The purpose is to calibrate the models $\mathcal{M}_3$ and $\mathcal{M}_4$ (Figure 2) where the south zone temperature (green zone Figure 1) is the output and two boundary conditions are considered, the outdoor temperature $T_o$ (°C) and the north zone temperature $T_n$ (°C). The vector of unknown parameters $\boldsymbol{\theta}$ for $\mathcal{M}_3$ and $\mathcal{M}_4$ are respectively given in Table 1 and Table 2. The process noise covariance is defined by $\boldsymbol{\Sigma}_\mathbf{w}(\boldsymbol{\theta}) = \text{diag}(\sigma_{w_{11}}^2 \quad \sigma_{w_{22}}^2 \quad \sigma_{w_{33}}^2)$ and the measurement noise variance by $\Sigma_v(\boldsymbol{\theta}) = \sigma_v^2$. The standard deviation $\sigma_{w_{44}}$ of the state $x_s$ in $\mathcal{M}_4$ has been fixed to $10^{-6}$ instead of putting an informative prior distribution.

This problem has been investigated by different authors who used relatively similar optimization strategies but different model structures. De Coninck et al. (2015) and Rehab & André (2015) used second order thermal models but with different structure and inputs. Himpe & Janssens (2015) used a model with four states which was correlated with the HVAC system and the solar radiations. They improved their model by scaling the system noise with respect to the heater and solar radiation signal. A validation data set has not been used to demonstrate the effectiveness of the model; only the improvement of the residuals is shown. With a zero order-hold assumption, the model is not able to understand the fast dynamic of the heating signal. Indeed, the time response of the electric heaters is estimated between 1 and 2 minutes, which is faster than the sampling time of the data (10 minutes). This issue is solved by considering that the inputs vary linearly between two samples (first order-hold).

Around 24 days of data are used, where the first 14 days are the identification data set (ROLBS sequence, Figure 4) and the last 10 days are the validation data set (Figure 5); the detail of the inputs and outputs is given in section 2.

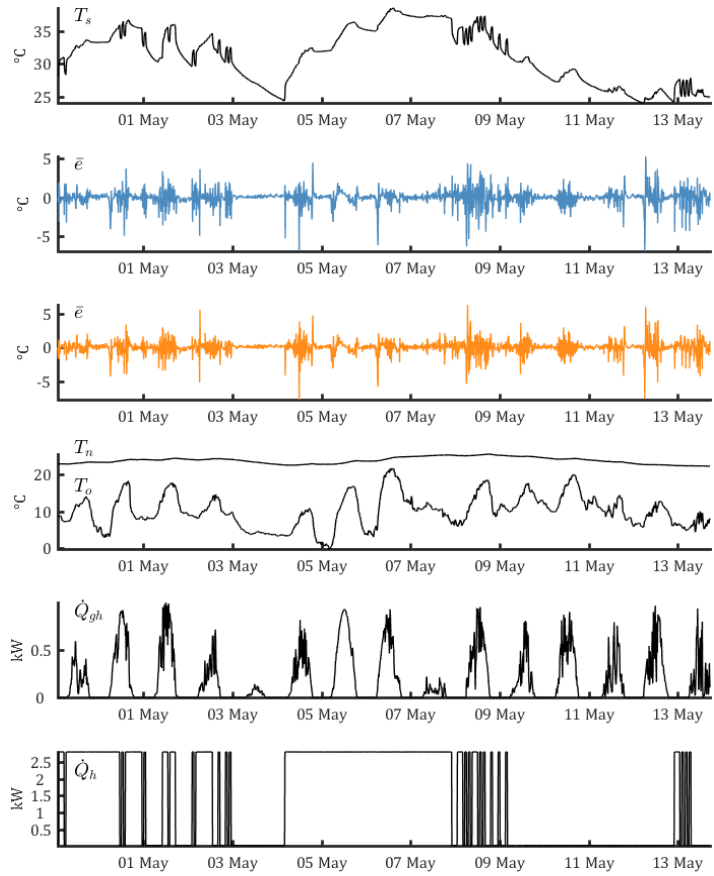

*Figure 4: Identification data set, output, standardized residuals of $\mathcal{M}_3$ (blue) and $\mathcal{M}_4$ (orange), and inputs*

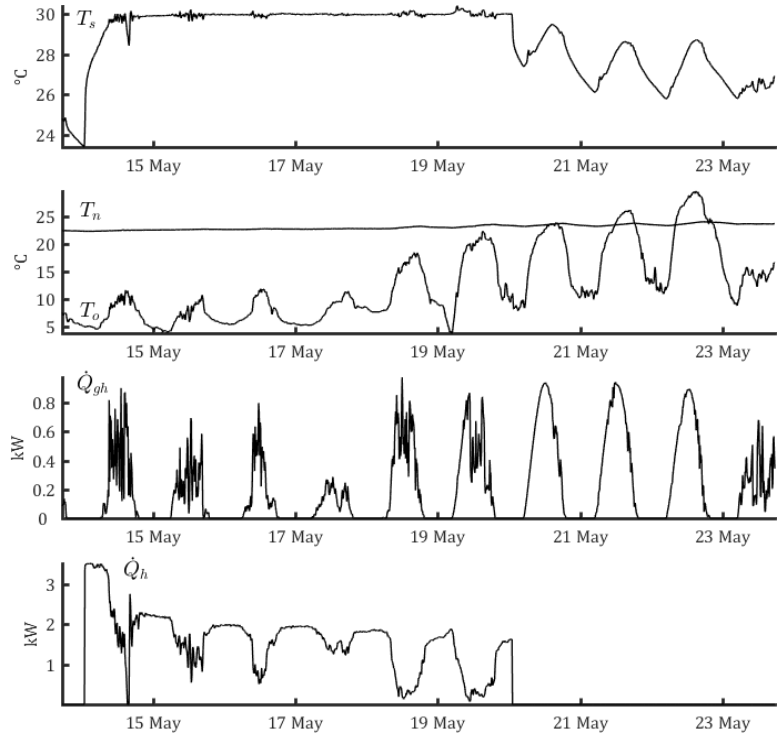

*Figure 5: Validation data set*

A unique step length in the proposal distribution has been used with $\boldsymbol{\Sigma} = \varepsilon \mathbf{I}_{Np}$, where $\varepsilon = 0.3$ has been selected such that it minimizes the IACT. This tuning gives an acceptance rate of approximately 30% for $\mathcal{M}_3$ and 25% for $\mathcal{M}_4$. The diagnosis procedure presented in section 3.2.5 has been applied with $M = 6$ Markov chains with initial parameter values randomly sampled from their respective prior distributions (Table 1 and Table 2); this is illustrated by the trace plot of the first thousand iterations in Figure 6. The first 500 samples of $N = 5500$ are discarded as burn-in for the model $\mathcal{M}_3$ whereas the first 1500 samples of $N = 6500$ are discarded for $\mathcal{M}_4$.

The chains at stationarity for both models are split in two to give 12 chains of 2500 samples which are used to quantify the mixing of the MH algorithm. The results are summarized in Table 1 and Table 1, with the worst values highlighted in red and the best in green. The worst potential scale reductions $\hat{R}$ are below the threshold of 1.1 and the worst effective samples sizes (ESS) are easily above $5 \times 12$. Consequently the $12 \times 2500$ samples can be used to approximate the posterior distributions of the parameters. The posterior distributions of the 6 simulated Markov chains are represented in Figure 7 for $\mathcal{M}_3$ and in Figure 8 for $\mathcal{M}_4$ with different colors; the black line is the global approximation of the posterior distributions using all samples.

Table 1: Prior distributions, posterior modes and diagnosis tests (worst: red and best: green) of $\mathcal{M}_3$

|  | Prior distributions | Posterior modes | $\hat{R}$ | ESS | Min/Max $\widehat{\mathrm{IACT}}$ | |
|---|---|---|---|---|---|---|
| $R_o$ | $\beta(2,2,10^{-4},2\cdot 10^{-1})$ | $5.53 \cdot 10^{-2}$ | 1.0100 | $1.22 \cdot 10^3$ | 18.86 | 29.47 |
| $R_i$ | $\beta(2,2,10^{-4},2\cdot 10^{-1})$ | $2.09 \cdot 10^{-3}$ | **1.0039** | $1.29 \cdot 10^3$ | 18.25 | 29.50 |
| $R_m$ | $\beta(2,2,10^{-4},10^{-1})$ | $2.31 \cdot 10^{-3}$ | 1.0045 | $1.25 \cdot 10^3$ | 17.81 | 30.81 |
| $R_z$ | $\beta(2,2,10^{-4},10^{-1})$ | $4.98 \cdot 10^{-3}$ | 1.0086 | $1.16 \cdot 10^3$ | 18.79 | 28.16 |
| $C_w/10^8$ | $\beta(2,2,10^{-3},5\cdot 10^{-1})$ | $3.12 \cdot 10^{-2}$ | 1.0049 | $1.31 \cdot 10^3$ | 19.46 | 27.42 |
| $C_i/10^8$ | $\beta(2,2,10^{-4},10^{-1})$ | $6.73 \cdot 10^{-3}$ | 1.0080 | $1.28 \cdot 10^3$ | 17.49 | 32.01 |
| $C_m/10^8$ | $\beta(2,2,10^{-2},5)$ | $1.38 \cdot 10^{-1}$ | 1.0061 | **$1.49 \cdot 10^3$** | 17.18 | 24.58 |
| $a_W$ | $\beta(2,2,10^{-1},5)$ | 1.06 | 1.0132 | $1.19 \cdot 10^3$ | 18.90 | 24.12 |
| $a_I$ | $\beta(2,2,10^{-1},5)$ | 1.24 | 1.0143 | $1.39 \cdot 10^3$ | 18.40 | **22.99** |
| $\sigma_{w_{11}}$ | $\mathcal{G}(2,0.03)$ | $1.02 \cdot 10^{-1}$ | 1.0065 | $1.26 \cdot 10^3$ | 19.06 | 30.88 |
| $\sigma_{w_{22}}$ | $\mathcal{G}(2,0.03)$ | $1.38 \cdot 10^{-2}$ | **1.0229** | **$5.28 \cdot 10^2$** | **36.29** | **54.90** |
| $\sigma_{w_{33}}$ | $\mathcal{G}(2,0.03)$ | $1.82 \cdot 10^{-2}$ | 1.0084 | $1.15 \cdot 10^3$ | 22.12 | 33.57 |
| $\sigma_v$ | $\mathcal{G}(2,0.03)$ | $1.69 \cdot 10^{-2}$ | 1.0179 | $6.26 \cdot 10^2$ | 29.39 | 50.03 |
| $x_{w_0}$ | $\beta(2,2,15,45)$ | 29.20 | 1.0146 | $1.15 \cdot 10^3$ | **16.43** | 24.49 |
| $x_{m_0}$ | $\beta(2,2,15,45)$ | 29.45 | 1.0099 | $1.33 \cdot 10^3$ | 18.09 | 27.53 |

Table 2: Prior distributions, posterior modes and diagnosis tests (worst: red and best: green) of $\mathcal{M}_4$

|  | Prior distributions | Posterior modes | $\hat{R}$ | ESS | Min/Max $\widehat{\mathrm{IACT}}$ |
|---|---|---|---|---|---|
| $R_o$ | $\beta(2,2,10^{-4},10^{-1})$ | $4.93 \cdot 10^{-2}$ | 1.0073 | $1.06 \cdot 10^3$ | [24.17  41.44] |
| $R_i$ | $\beta(2,2,10^{-4},10^{-1})$ | $1.83 \cdot 10^{-3}$ | 1.0127 | $7.26 \cdot 10^2$ | [30.34  58.48] |
| $R_m$ | $\beta(2,2,10^{-4},10^{-1})$ | $2.13 \cdot 10^{-3}$ | 1.0136 | $6.04 \cdot 10^2$ | [34.27  66.75] |
| $R_s$ | $\beta(2,2,10^{-3},5\cdot 10^{-2})$ | $5.35 \cdot 10^{-3}$ | 1.0099 | $7.41 \cdot 10^2$ | [31.15  53.36] |
| $R_z$ | $\beta(2,2,10^{-4},10^{-1})$ | $5.27 \cdot 10^{-3}$ | 1.0094 | $9.89 \cdot 10^2$ | [24.93  35.84] |

| | | | | | | |
|---|---|---|---|---|---|---|
| $C_w/10^8$ | $\beta(2,2,10^{-3},5\cdot 10^{-1})$ | $2.52\cdot 10^{-2}$ | 1.0176 | $5.23\cdot 10^2$ | [37.62 | 69.45] |
| $C_i/10^8$ | $\beta(2,2,10^{-4},10^{-2})$ | $5.27\cdot 10^{-3}$ | 1.0182 | $5.69\cdot 10^2$ | [35.82 | 89.49] |
| $C_m/10^8$ | $\beta(2,2,10^{-2},5)$ | $1.46\cdot 10^{-1}$ | 1.0083 | $1.13\cdot 10^3$ | [23.32 | 32.37] |
| $C_s/10^8$ | $\beta(2,2,10^{-5},10^{-3})$ | $5.81\cdot 10^{-5}$ | 1.0090 | $6.15\cdot 10^2$ | [37.72 | 83.21] |
| $a_W$ | $\beta(2,2,10^{-1},5)$ | $9.07\cdot 10^{-1}$ | 1.0181 | $6.33\cdot 10^2$ | [36.30 | 69.06] |
| $a_I$ | $\beta(2,2,10^{-1},5)$ | 1.17 | 1.0126 | $\mathbf{1.16\cdot 10^3}$ | [22.65 | **28.41**] |
| $\sigma_{w_{11}}$ | $\mathcal{G}(2,0.03)$ | $9.87\cdot 10^{-2}$ | 1.0095 | $9.69\cdot 10^2$ | [24.21 | 41.21] |
| $\sigma_{w_{22}}$ | $\mathcal{G}(2,0.03)$ | $2.61\cdot 10^{-2}$ | 1.0248 | $\mathbf{4.08\cdot 10^2}$ | [**43.92** | 86.54] |
| $\sigma_{w_{33}}$ | $\mathcal{G}(2,0.03)$ | $3.18\cdot 10^{-2}$ | **1.0285** | $4.35\cdot 10^2$ | [41.75 | **110.54**] |
| $\sigma_v$ | $\mathcal{G}(2,0.03)$ | $1.15\cdot 10^{-2}$ | 1.0233 | $4.31\cdot 10^2$ | [42.91 | 81.03] |
| $x_{w_0}$ | $\beta(2,2,15,45)$ | 28.83 | 1.0124 | $1.02\cdot 10^3$ | [27.08 | 35.02] |
| $x_{m_0}$ | $\beta(2,2,15,45)$ | 28.31 | **1.0070** | $1.08\cdot 10^3$ | [24.45 | 34.56] |
| $x_{i_0}$ | $\beta(2,2,15,45)$ | 29.64 | 1.0099 | $9.49\cdot 10^2$ | [**21.95** | 44.21] |

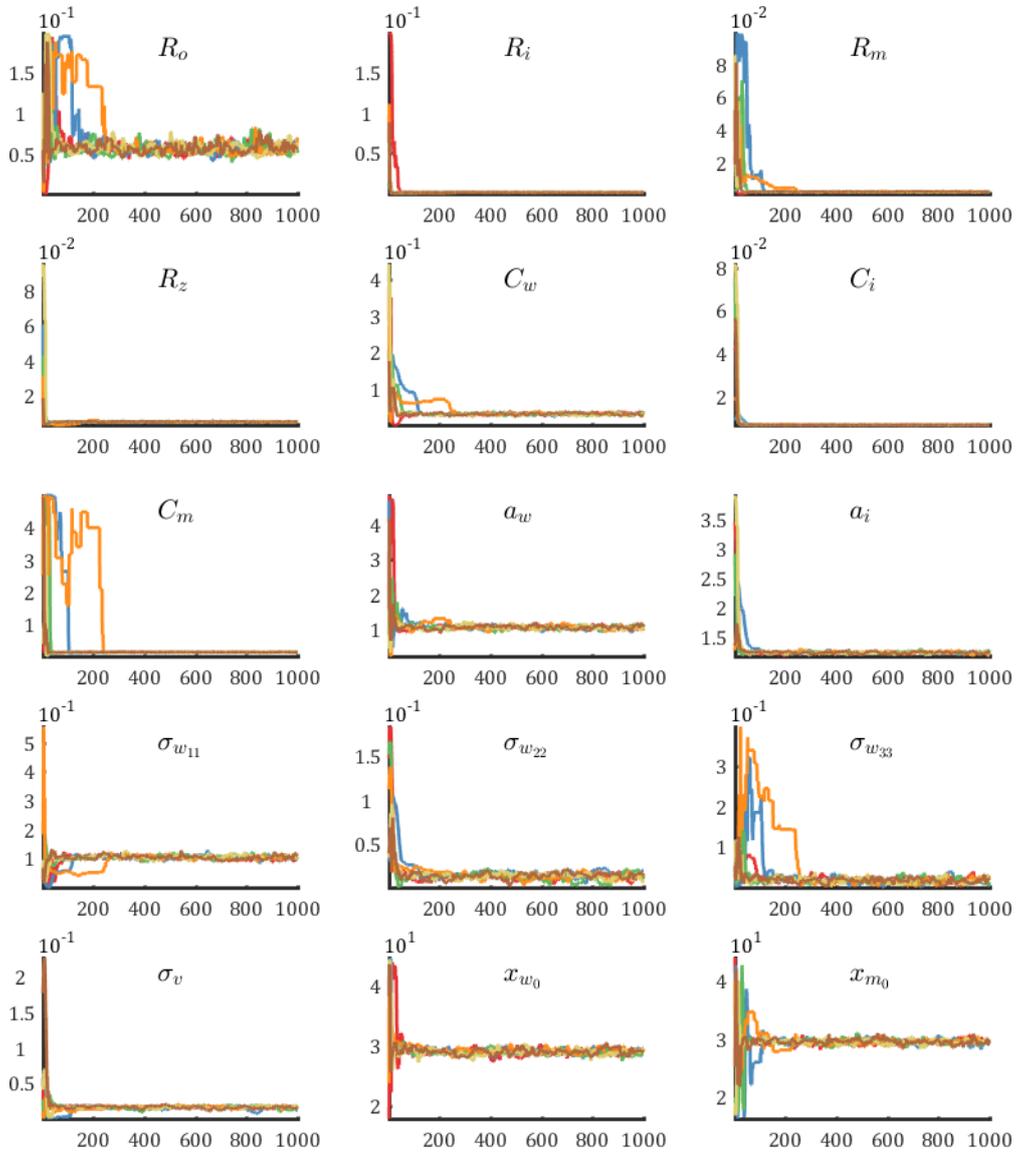

*Figure 6: Trace plot of the 6 Markov chains in the first thousand iterations ($\mathcal{M}_3$)*

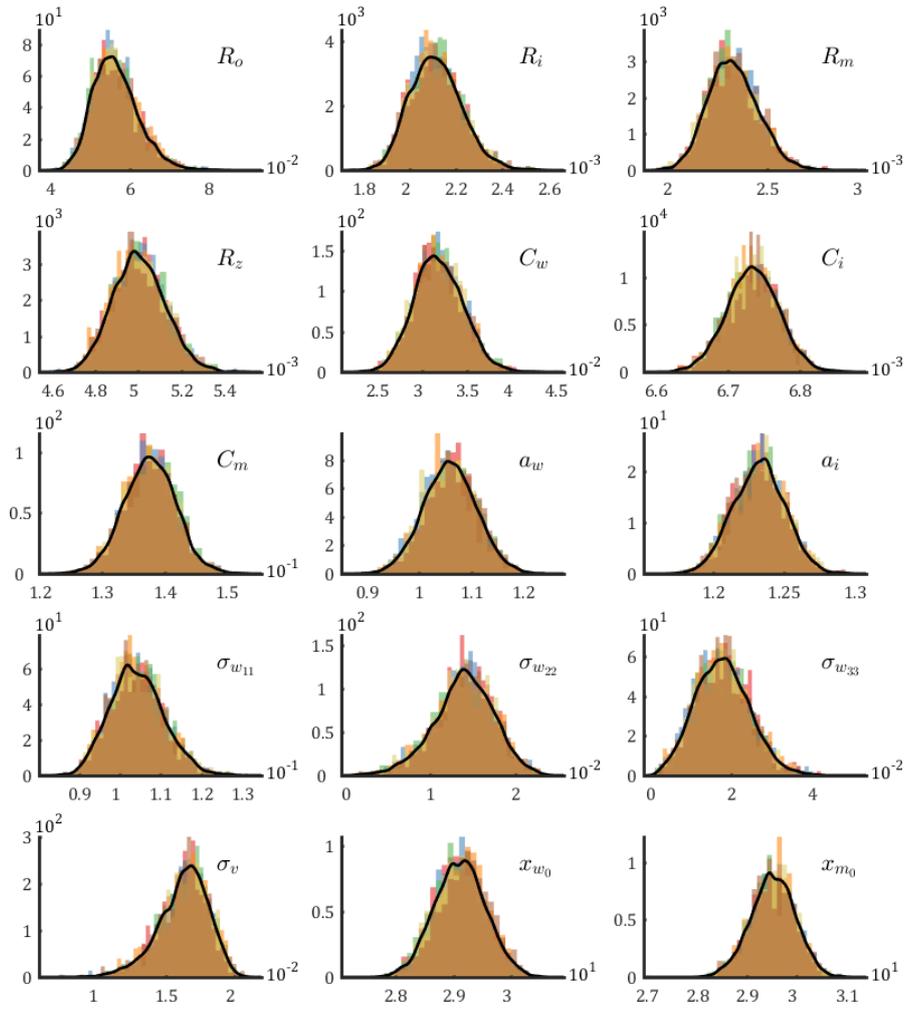

*Figure 7: Posterior distribution of $\mathcal{M}_3$ from the 6 Markov chains*

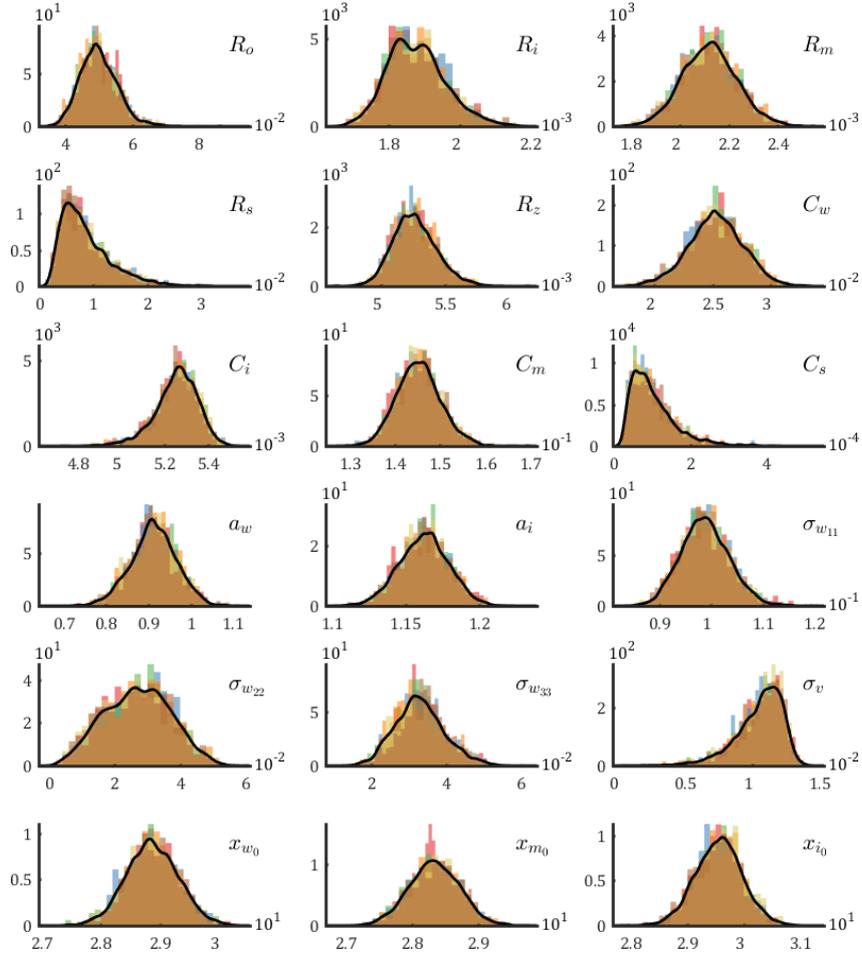

*Figure 8: Posterior distribution of $\mathcal{M}_4$ from the 6 Markov chains*

The reliability of the calibrated models is tested by residual analysis. The least correlated standardized residuals $\bar{\mathbf{e}}$ (13) are shown in Figure 4, the ACF, CCF and CP of models $\mathcal{M}_3$ and $\mathcal{M}_4$ are shown respectively in Figure 9 and in Figure 10. The red lines delimit the 95% confidence intervals and the lags are the number sample shifts between the two signals. Hence, to validate the hypothesis, 5% of the lags must not cross these limits. For both models, the inputs are not correlated with the standardized residuals which means that the models are able to explain all input-output relationships. However, the white noise hypothesis is rejected for the model $\mathcal{M}_3$; the highest standardized residual values (Figure 4) are correlated with the switches of the heating signal and the solar radiations.

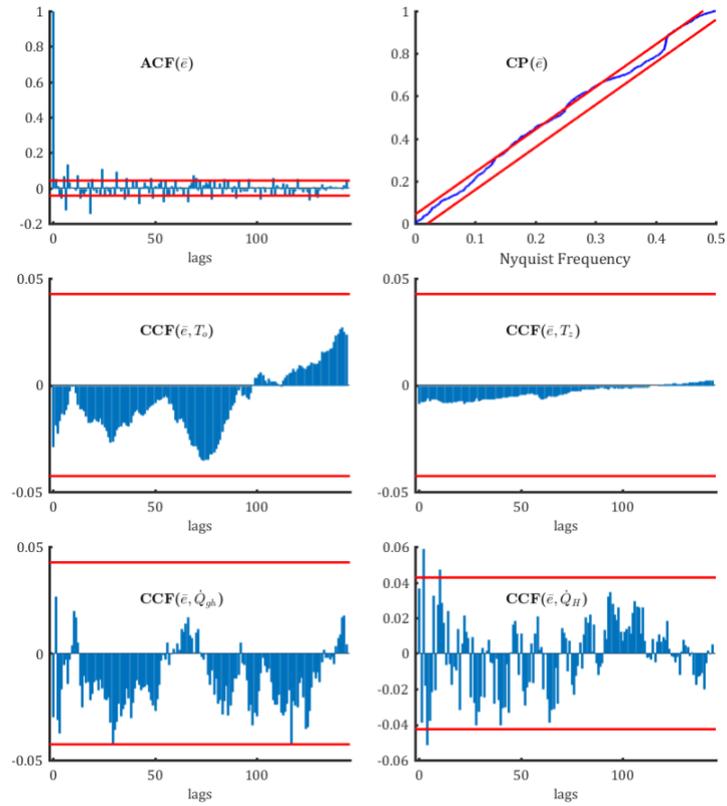

*Figure 9: ACF, CP and CCF of the standardized residuals, $\mathcal{M}_3$*

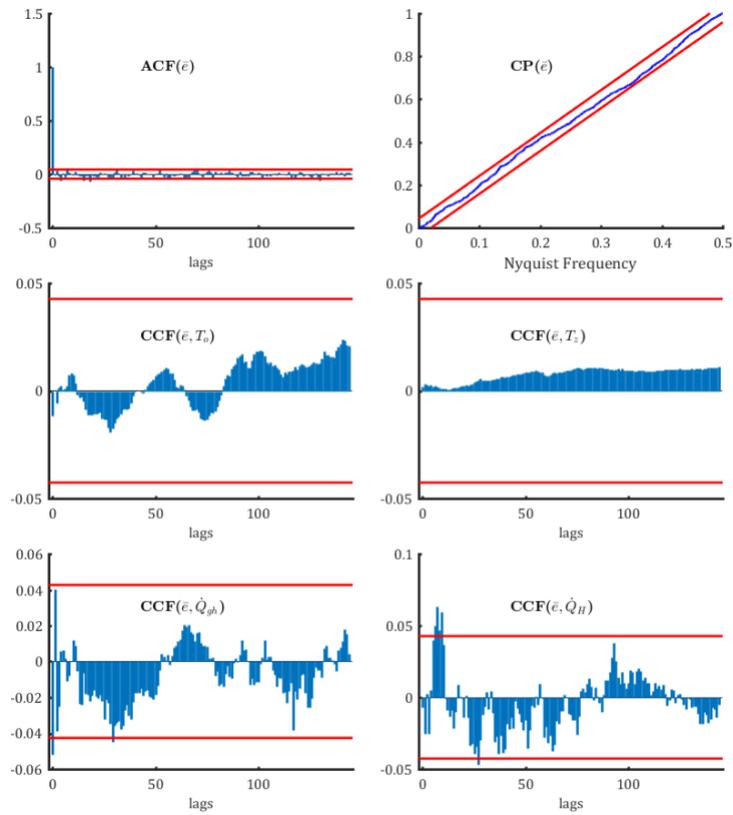

*Figure 10: ACF, CP and CCF of the standardized residuals, $\mathcal{M}_4$*

The reliability of the model is also tested by comparing the measured south zone temperature of the validation data set with the simulated output. A clear advantage of Bayesian estimation is that it is possible to simulate directly from the posterior distribution, which gives a simulated output with all the uncertainties. This is very useful for model predictive control; the weather forecast is introduced in the estimated model to predict the indoor temperature. Afterwards, the trade-off between comfort and energy saving is chosen by taking either the lowest temperature prediction, such that the HVAC system is sure to maintain the comfort or by taking the highest predicted temperature such that the HVAC system uses the minimal amount of energy. The simulation from the posterior distribution is plotted in Figure 11. The measured south zone temperature is always included in the simulated output distribution for both models, but the dispersion for the model $\mathcal{M}_4$ is more important.

In order to select the most appropriate model, the performances of both models (section 3.4) are summarized in Table 3. For the identification data set, the model $\mathcal{M}_4$ should be accepted against $\mathcal{M}_3$; the AIC, BIC and WAIC are smaller for $\mathcal{M}_4$ than for $\mathcal{M}_3$, and the $p_{Value}$ of the LRT confirms this choice. However, the criteria evaluated on the validation data set indicate that the model $\mathcal{M}_3$ should be preferred; the log-likelihood of the model $\mathcal{M}_3$ is higher and less dispersed than the log-likelihood of the model $\mathcal{M}_4$, as shown in Figure 11.

*Table 3: Performance criteria*

|  | **Identification data set** | | **Validation data set** | |
| --- | --- | --- | --- | --- |
|  | $\mathcal{M}_3$ | $\mathcal{M}_4$ | $\mathcal{M}_3$ | $\mathcal{M}_4$ |
| AIC | $-7.59 \cdot 10^3$ | $-8.02 \cdot 10^3$ | $-5.26 \cdot 10^3$ | $-5.21 \cdot 10^3$ |
| BIC | $-7.51 \cdot 10^3$ | $-7.91 \cdot 10^3$ | $-5.19 \cdot 10^3$ | $-5.11 \cdot 10^3$ |
| LRT $p_{Value}$ | 0 | | 1 | |
| WAIC | $-7.58 \cdot 10^3$ | $-8.01 \cdot 10^3$ | $-5.27 \cdot 10^3$ | $-5.23 \cdot 10^3$ |

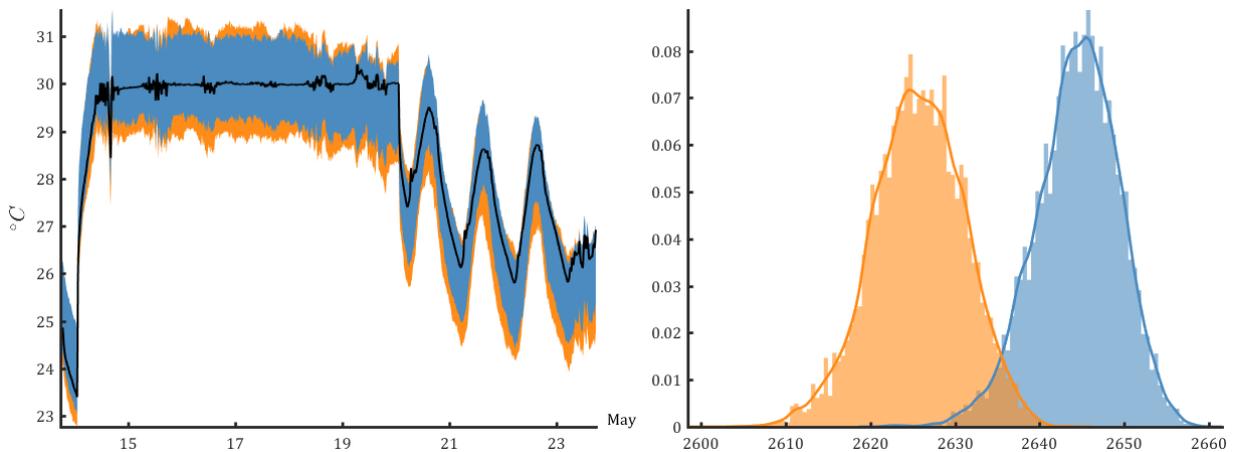

*Figure 11: Left: measured (black) and simulated south zone temperature with the validation data set; right: log-likelihoods for the validation data set, ($\mathcal{M}_3$: blue, $\mathcal{M}_4$: orange)*

The performance gap between both models is significant for the identification data set in comparison to the validation data set. In the identification data set, the ROLBS introduces an unconventional dynamic which is not representative of the intended use of the building. A more complex model is required to fit the fast variations of the south zone air temperature, but these fast variations are not present in a conventional use and therefore the extra complexity of the model $\mathcal{M}_4$ is not required. As mentioned in section 3.4, using the same information from the data for calibration and for selection may be misleading (Gelman, Hwang, et al. 2014). In this case, it is also illustrated by the whiteness improvement of the residuals for the model $\mathcal{M}_4$. It can be concluded that the model $\mathcal{M}_3$ is more representative of the south zone and consequently, only the model $\mathcal{M}_3$ is considered in the following of the paper.

### 4.2. Physical interpretation of the results

The posterior distributions are compared against the building characteristics which are available in Strachan et al. (2016). The envelope thermal resistance is given by

$$\bar{R}_w = \left(\sum_{j=1}^{3} U_j S_j + U_{wi} S_{wi}\right)^{-1} = 4.47 \cdot 10^{-2} \text{ K/W} \tag{43}$$

where $U_j$ and $S_j$ are the U-values and surfaces of the different walls (south, east and west) and $U_{wi}$ and $S_{wi}$ are the U-value and surfaces of the windows.

The envelope thermal resistance is estimated with the posterior distributions of $R_o$ and $R_i$ such that

$$R_w = R_o + R_i \in [4.11 \cdot 10^{-2} \quad 9.38 \cdot 10^{-2}] \tag{43.a}$$

The resistance $\bar{R}_w$ belongs to the estimated interval (43.a), but this interval is large which means that the uncertainties are important.

The thermal resistance between the south and the north zone is given by

$$\bar{R}_z = (U_{z1} S_{z1} + U_{z2} S_{z2} + 3 U_{door} S_{door})^{-1} = 1.53 \cdot 10^{-2} \text{ K/W} \tag{44}$$

where $U_{z1}$ and $S_{z1}$ are the U-value and the surface of the north wall of the living room, $U_{z2}$ and $S_{z2}$ are the U-value and the surface of the north wall of the bathroom and the corridor, and $U_{door}$ and $S_{door}$ are the U-value and the surface of the doors.

The estimated range from the posterior distribution of $R_z$ is $[4.59 \cdot 10^{-3} \quad 5.52 \cdot 10^{-3}]$ which is far smaller than the value computed in (44). This gap could mean that the infiltration between the two zones are significant.

The envelope thermal capacity is defined by

$$\bar{C}_w = \sum_{j=1}^{3} C_j S_j = 1.30 \cdot 10^7 \text{ J/K} \tag{45}$$

with $C_j$ the heat capacity. The thermal capacity of the windows is negligible as compared to $C_w$.

The estimated posterior distribution of $C_w$ covers the following range of values $[2.21 \cdot 10^6 \quad 4.47 \cdot 10^6]$, which is more than half less than the expected value (45).

The thermal capacity of the medium $C_m$ consists of the inner walls of the south zone but also of parts of the ceiling and ground floor, such as

$$\bar{C}_m = C_{iw_1} + C_{iw_2} + C_{ground} + C_{ceiling} = 8.00 \cdot 10^7 \text{ J/K} \tag{46}$$

where the subscripts $iw_1$ and $iw_2$ denote respectively the east wall of the living room and the other light walls.

The estimated range for $C_m$ is $[1.22 \cdot 10^7 \quad 1.54 \cdot 10^7]$ which is the expected order of magnitude since (46) is overestimated by taking into account all the volume of the ground floor and the ceiling.

The thermal capacity of the indoor air is simply given by

$$\bar{C}_i = \rho_a c_a V = 1.79 \cdot 10^5 \text{ J/K} \tag{47}$$

with $c_a$ and $\rho_a$ the specific heat and the density of the air, and $V$ the volume of the south zone.

In this case as well, the estimated posterior distribution of $C_i$ is consistent with the order of magnitude of the expected value, where $C_i \in [6.60 \cdot 10^5 \quad 6.90 \cdot 10^5]$.

Determining prior knowledge on the convective resistance $R_m$ between the indoor air and the medium is not an easy task and it is not of main interest in this study. The parameters $a_I$ and $a_W$ are interpreted as effective areas because $\dot{Q}_{gh}$ is measured on a horizontal surface. Nevertheless, it is interesting to see that $a_I$ is superior to $a_W$ which shows the importance of direct solar radiations into the south zone.

The time constants $\boldsymbol{\tau}$ of the continuous system (4) are computed by

$$\boldsymbol{\tau} = -\frac{1}{\boldsymbol{\lambda}} \tag{48}$$

where $\boldsymbol{\lambda}$ are the eigenvalues of the state matrix and $\boldsymbol{\tau}, \boldsymbol{\lambda} \in \mathbb{R}$ ,.

The time constants of the model $\mathcal{M}_3$, given in Table 4, are consistent in the range of the fast dynamics of the air and the slow accumulation of energy in the medium.

The performances of the second-order MH are compared to a ML estimation in the next section. Furthermore, the regularization effect of the prior distribution is illustrated by identifiability analysis.

Table 4: Time constants of model $\mathcal{M}_3$

| Time constant [hours] | $\mathcal{M}_3$ |
|---|---|
| $\tau_1$ | $[1.57 \cdot 10^{-1} \quad 1.66 \cdot 10^{-1}]$ |
| $\tau_2$ | $[2.19 \quad 3.75]$ |
| $\tau_3$ | $[25.84 \quad 34.34]$ |

## 4.3. Performance comparison with maximum likelihood estimation

Table 5: ML estimation with the same random initial parameters as the MH algorithm, the values in bold represent the parameters closed to their boundaries

|  | MLE 1 | MLE 2 | MLE 3 | MLE 4 | MLE 5 | MLE 6 |
|---|---|---|---|---|---|---|
| $R_o$ | $\mathbf{1.98 \cdot 10^{-1}}$ | $\mathbf{1.02 \cdot 10^{-4}}$ | $\mathbf{1.98 \cdot 10^{-1}}$ | $\mathbf{1.98 \cdot 10^{-1}}$ | $1.62 \cdot 10^{-1}$ | $\mathbf{1.98 \cdot 10^{-1}}$ |
| $R_i$ | $1.40 \cdot 10^{-3}$ | $4.78 \cdot 10^{-2}$ | $1.40 \cdot 10^{-3}$ | $1.18 \cdot 10^{-3}$ | $1.35 \cdot 10^{-3}$ | $1.42 \cdot 10^{-3}$ |
| $R_m$ | $2.24 \cdot 10^{-2}$ | $1.12 \cdot 10^{-3}$ | $2.19 \cdot 10^{-2}$ | $3.32 \cdot 10^{-3}$ | $2.91 \cdot 10^{-3}$ | $2.08 \cdot 10^{-2}$ |
| $R_z$ | $2.94 \cdot 10^{-3}$ | $5.76 \cdot 10^{-3}$ | $2.92 \cdot 10^{-3}$ | $4.00 \cdot 10^{-3}$ | $\mathbf{8.61 \cdot 10^{-2}}$ | $2.87 \cdot 10^{-3}$ |
| $C_w/10^8$ | $6.50 \cdot 10^{-2}$ | $\mathbf{1.01 \cdot 10^{-3}}$ | $6.46 \cdot 10^{-2}$ | $1.19 \cdot 10^{-1}$ | $6.94 \cdot 10^{-2}$ | $6.39 \cdot 10^{-2}$ |
| $C_i/10^8$ | $6.75 \cdot 10^{-3}$ | $6.86 \cdot 10^{-3}$ | $6.76 \cdot 10^{-3}$ | $6.72 \cdot 10^{-3}$ | $6.74 \cdot 10^{-3}$ | $6.81 \cdot 10^{-3}$ |
| $C_m/10^8$ | $3.03 \cdot 10^{-1}$ | $1.24 \cdot 10^{-1}$ | $3.12 \cdot 10^{-1}$ | $\mathbf{1.33 \cdot 10^{-2}}$ | 2.3 | $3.29 \cdot 10^{-1}$ |
| $a_W$ | 1.01 | 3.15 | 1.02 | 1.74 | 1.08 | 1.03 |
| $a_I$ | 1.23 | 1.27 | 1.24 | 1.27 | 1.23 | 1.25 |
| $\sigma_{w_{11}}$ | $5.83 \cdot 10^{-2}$ | $1.14 \cdot 10^{-2}$ | $5.50 \cdot 10^{-2}$ | $8.71 \cdot 10^{-2}$ | $5.09 \cdot 10^{-2}$ | $4.67 \cdot 10^{-2}$ |
| $\sigma_{w_{22}}$ | $3.28 \cdot 10^{-5}$ | $\mathbf{1.15 \cdot 10^{-8}}$ | $1.30 \cdot 10^{-2}$ | $1.32 \cdot 10^{-2}$ | $2.90 \cdot 10^{-5}$ | $2.51 \cdot 10^{-2}$ |
| $\sigma_{w_{33}}$ | $6.45 \cdot 10^{-1}$ | $6.58 \cdot 10^{-2}$ | $6.41 \cdot 10^{-1}$ | $1.57 \cdot 10^{-5}$ | $9.79 \cdot 10^{-2}$ | $6.38 \cdot 10^{-1}$ |
| $\sigma_v$ | $1.90 \cdot 10^{-2}$ | $1.70 \cdot 10^{-2}$ | $1.63 \cdot 10^{-2}$ | $1.63 \cdot 10^{-2}$ | $1.90 \cdot 10^{-2}$ | $\mathbf{1.38 \cdot 10^{-8}}$ |
| $x_{w_0}$ | 29.60 | 36.68 | 29.59 | 29.54 | 29.70 | 29.60 |
| $x_{m_0}$ | $\mathbf{44.47}$ | 29.61 | $\mathbf{44.50}$ | 29.95 | 25.00 | $\mathbf{44.57}$ |

The performance of the second-order MH algorithm is now tested against a ML estimation with a quasi-Newton optimization. The log-likelihood and its gradient are supplied to the unconstrained MATLAB's function *fminunc*[1] (optimality tolerance and step tolerance fixed to $10^{-8}$) which use a *BFGS* approximation of the Hessian. The same parameter transformations are used except for the standard deviations; they are bounded between $10^{-8}$ and 5 because it has been observed that with the logarithm transformation (33), the standard deviations can be too close to zero which cause numerical instabilities. The penalty function given by Kristensen & Madsen (2003) is used to repulse the parameters near the bounds. The ML estimation is repeated 6 times with the same initial parameters as for MH algorithm; the ML parameter estimates are given in Table 5. The results are highly dependent on the initial conditions for most of the parameters; some of them are different of several orders of magnitude. It can be concluded in this case that the second-order MH algorithm has a better global convergence than the ML optimization.

---

[1] *https://fr.mathworks.com/help/optim/ug/fminunc.html*

A closer look is given to the parameter $\sigma_{w_{22}}$ which seems to be unidentifiable compared to the parameter $C_i$. The profiles of the log-likelihood and log-posterior are plotted in Figure 12. These profiles are obtained by maximizing the log-likelihood and the log-posterior with respect to all parameters except $\sigma_{w_{22}}$ and $C_i$, which are fixed for each optimization to different values (x-axis, Figure 12). Confidence intervals for the profile likelihood can be computed by (Madsen & Thyregod 2012)

$$\ln p(\mathbf{y}_{1:k}|\boldsymbol{\theta}_{\sim i}) - \ln p(\mathbf{y}_{1:k}|\boldsymbol{\theta}) > -\frac{1}{2}\chi^2_{1-\alpha}(p) \tag{49}$$

where $\ln p(\mathbf{y}_{1:k}|\boldsymbol{\theta}_{\sim i})$ is the maximum log-likelihood with respect to all parameters except $\theta_i$, $\ln p(\mathbf{y}_{1:k}|\boldsymbol{\theta})$ is the maximum log-likelihood and $\chi^2_{1-\alpha}(p)$ is a chi squared distribution with $p$ degrees of freedom and confidence level $\alpha$.

The 95% confidence intervals given by $-\frac{1}{2}\chi^2_{0.95}(1) = -1.92$ are represented in Figure 12 by the dotted black lines. The profile of the log-posterior is similarly computed by $\ln p(\boldsymbol{\theta}_{\sim i}|\mathbf{y}_{1:k}) - \ln p(\boldsymbol{\theta}|\mathbf{y}_{1:k})$. These profiles are computed around the posterior modes given in Table 1; the idea is to visualize the quantity of information given by the data and the regularization brought by the prior distribution. The profile of the log-likelihood of $\sigma_{w_{22}}$ is asymmetric and almost flat around its maximum, which means that any value on the flat region has negligible effects on the log-likelihood; it explains the dispersion of the ML estimation in Table 5. The profile of the log-posterior shows how the prior distribution ($\mathcal{G}(2, 0.03)$, Figure 3) increases the curvature, especially towards zero, and regularizes the identifiability problem. For the parameter $C_i$, the information from the data overweight the prior distribution which means that the prior distribution has only a small effect on the posterior distribution.

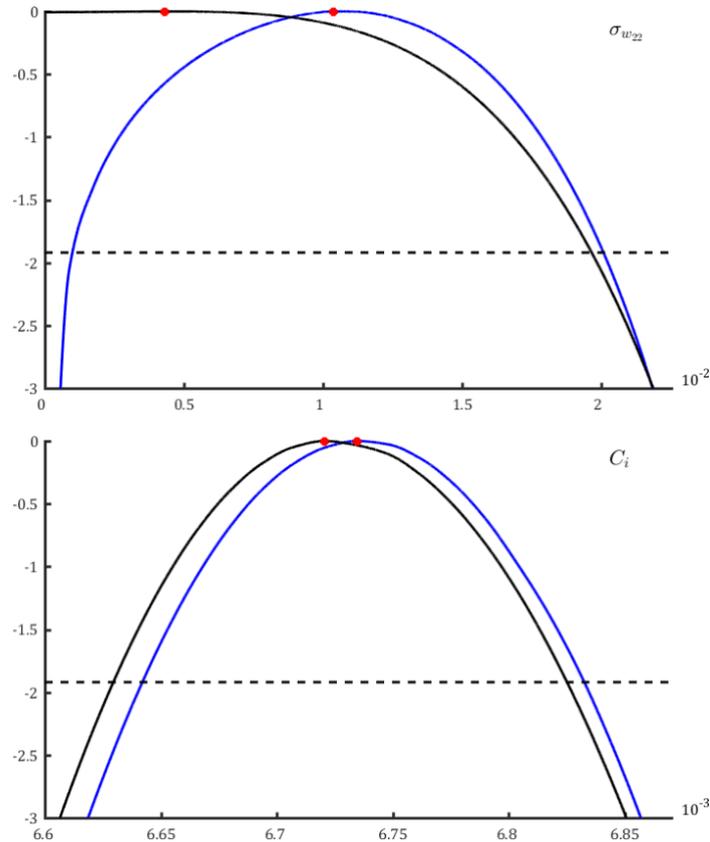

*Figure 12: Profile log-likelihood (black) and profile log-posterior (blue). The dashed lines represent the 95% confidence intervals of the log-likelihood and the red dots the respective maximums of the curves*

5. Conclusion

The estimation of building energy demand and building energy performance are possible through experimental calibration of dynamic thermal models. Making decisions or predictions from the calibrated model requires to take into account all the uncertainties of the estimates; Bayesian calibration fits this purpose by estimating the posterior distributions of the parameters.

This paper compares the three phases of an experimental calibration (selection, calibration and validation), from a Bayesian and a frequentist point of view. More specifically, proposed improvements on the Metropolis-Hastings algorithm, using gradient and Hessian information (second-order Metropolis-Hastings) are presented. It is shown that the gradient of linear and Gaussian model can be computed exactly by a robust square root version of the Kalman filter, and a Hessian estimate is proposed with low extra computational burden. A combination of change of variable and prior distribution is also proposed, which allows to constrain the parameters in a physical range. These improvements on the Metropolis-Hastings facilitate considerably the tuning of the algorithm: only a step length and the prior distributions have to be specified.

Two models of respectively 15 and 18 unknown parameters have been easily calibrated with the improved Metropolis-Hastings algorithms where a unique step-length has been used, which illustrates the gain of this method over a classical Metropolis-Hastings with random walk. Furthermore, it is shown that the second-order Metropolis-Hastings algorithm has a better robustness against the initial conditions than a maximum likelihood estimation with a quasi-Newton algorithm, and it is illustrated through an identifiability analysis, that the prior distributions act as regularization when the data are not informative enough.

It is highlighted that model selection criteria should be computed on a different data set than the one used for the calibration to avoid a biased selection. Indeed, in this experiment the unconventional excitation generated by the heaters implies that a more complex model should be selected, but this extra complexity is not required for a more conventional use of the HVAC system.

1. Acknowledgements

This work was financially supported by BPI France in the FUI Project COMETE. Thanks you to Dr. Paul Strachan for making the Twin houses data available online; to the Dynastee network, particularly to Henrik Madsen, Peder Bacher and Rune Juhl for their statistical guidelines; as well as Dr. Johan Dahlin and Dr. Maria Kulikova for their time and expertise.